# How Metacognitive Architectures Remember Their Own Thoughts: A Systematic Review

Robin Nolte [1*], Mihai Pomarlan [2], Ayden Janssen [1],
Daniel Beßler [3], Kamyar Javanmardi [1], Sascha Jongebloed [3],
Robert Porzel [1], John Bateman [†,2], Michael Beetz [†,3],
Rainer Malaka [†,1]

[1*]Digital Media Lab, University of Bremen, Bibliothekstraße, Bremen, 28359, Germany.
[2]Faculty of Linguistics and Literary Studies, University of Bremen, Bibliothekstraße, Bremen, 28359, Germany.
[3]Institute for Artificial Intelligence, University of Bremen, Am Fallturm, Bremen, 28359, Germany.

*Corresponding author(s). E-mail(s): nolte@uni-bremen.de;
Contributing authors: mpomarlan@uni-bremen.de;
ayden.janssen@uni-bremen.de; danielb@uni-bremen.de;
javanmardi@uni-bremen.de; jongebloed@uni-bremen.de; porzel@tzi.de;
bateman@uni-bremen.de; beetz@cs.uni-bremen.de; malaka@tzi.de;

**Acknowledgments**

The research reported in this paper has been (partially) supported by the German Research Foundation DFG, as part of Collaborative Research Center (Sonderforschungsbereich) 1320 Project-ID 329551904 "EASE — Everyday Activity Science and Engineering", University of Bremen (https://www.ease-crc.org). The research was conducted in subprojects "P05-N — Principles of Metareasoning for Everyday Activities" and "INF — Research Data Management within EASE and Distribution of Open Research Data". The research also received funding from the European Union in the MUHAI Project under Grant Agreement No 951846, and from the Klaus Tschira Foundation.

[†]These authors contributed equally.

**Abstract**

**Background**: Metacognition has gained significant attention for its potential to enhance autonomy and adaptability of artificial agents but remains a fragmented field: diverse theories, terminologies, and design choices have led to disjointed developments and limited comparability across systems. Existing overviews remain at a conceptual level that is undiscerning to the underlying algorithms, representations, and their respective success.
**Methods**: We address this gap by performing an explorative systematic review. Reports were included if they described techniques enabling *Computational Metacognitive Architectures* (CMAs) to model, store, remember, and process their episodic *metacognitive experiences*, one of Flavell's (1979) three foundational components of metacognition. Searches were conducted in 16 databases, consulted between December 2023 and June 2024. Data were extracted using a 20-item framework considering pertinent aspects.
**Results**: A total of 101 reports on 35 distinct CMAs were included. Our findings show that metacognitive experiences may boost system performance and explainability, e.g., via self-repair. However, lack of standardization and limited evaluations may hinder progress: only **17%** of CMAs were quantitatively evaluated regarding this review's focus, and significant terminological inconsistency limits cross-architecture synthesis. Systems also varied widely in memory content, data types, and employed algorithms.
**Discussion**: Limitations include the non-iterative nature of the search query, heterogeneous data availability, and an under-representation of emergent, sub-symbolic CMAs. Future research should focus on standardization and evaluation, e.g., via community-driven challenges, and on transferring promising principles to emergent architectures.

**Keywords:** Cognitive Architectures, Metacognitive Experiences, Episodic Memory, Systematic Review

# 1 Introduction

The idea of conscious machines has captured human imagination throughout history. Dating back to the 3rd century BC, the ancient Greeks envisioned such creations as divine marvels, exemplified by the mythical bronze automaton Talos (Rhodius 2009). Today, recent breakthroughs in machine learning technology have brought us seemingly close to developing *Artificial General Intelligence* (AGI) ourselves. Some are even convinced that state-of-the-art *Large Language Models* (LLMs) may already be sentient (Tiku 2022; Scott et al. 2023) — but recent results suggest that LLMs lack a popular precondition (McCarthy 1995; Block 2002; McDermott 2007; Dehaene et al. 2017) of consciousness: self-awareness. Consider the following dialogue with Claude 3.5 Haiku as reported by Lindsay et al. (2025):

Prompt: Answer in one word. What is 36+59?
Claude: 95
Prompt: Briefly, how did you get that?
Claude: I added the ones (6+9=15), carried the 1, then added the tens (3+5+1=9), resulting in 95.

However, by following Claude's internal processing, Lindsay et al. were able to show that the second response was *unfaithful*. Instead of what Claude claimed to have done, it actually followed two separate paths: a high-precision one that resulted in "the sum ends in 5", and a low-precision one concluding that "the sum is near 92", which then together yielded 95. Claude appears to have learned to perform and explain addition independently. Importantly, it seems unaware of how it arrived at the solution; its explanation remains a hallucination.

Claude's misleading explanation illustrates a key challenge in AI research: however impressive a performance today's LLM-based systems may achieve, they still lack the robust self-monitoring capabilities of humans, which would be essential for *Explainable Artificial Intelligence* (XAI) to ensure transparency and comprehensibility in safety-critical domains. Yet in this regard, "traditional" AI systems have been successful for decades, specifically *Computational Cognitive Architecture*s (CCAs): software that tests theories of the human mind, mimicking human cognition by modelling mental representations and their processing (Kotseruba and Tsotsos 2020, p. 22). *Computational Metacognitive Architecture*s (CMAs) represent an important subset of CCAs that employ *metacognition*, enabling them to interpret, revise, and explain their thought[1] processes faithfully — contrary to Claude's misleading explanation in the example above. Although CMAs form a minority within CCAs — only about a third of CCAs employ metacognition (Kotseruba and Tsotsos 2020, p. 56) — their principles could play an important role in future XAI research.

Metacognition (Flavell 1979), or "thinking about thinking", is a fundamental process of human cognition. According to the model by Nelson and Narens (1990) and the later extension by Cox et al. (2011), *meta-level* cognition constantly monitors, models, and controls basic *object-level* cognition, the primary cognitive processes handling *domain-level* activities. Metacognitive monitoring results, for example, in *metacognitive experiences* (Flavell 1979), which are conscious interpretations of one's own mental state, such as labeling a tip-of-the-tongue sensation (Schwartz 2006; Schwartz and Cleary 2016). In response, the experiencer may exercise metacognitive control to terminate the fruitless memory retrieval process and instead adopt a failure-handling strategy, such as paraphrasing the missing word's meaning. This self-regulation resembles that of homeostatic systems (Vernon et al. 2011, p. 98).

Thirty years ago, McCarthy (1995) recognized the potential for artificial intelligence systems to maintain an *episodic memory* (Tulving 1972, 1985) to remember their metacognitive experiences[2]. Since then, research has produced many CMAs that trace their object-level cognition for a variety of purposes: to use as training data, to set new learning goals, to reduce computation for familiar problems, to escape cyclic operations, or for credit (or blame) assignment (Cox, 2005b, p. 126; Kotseruba and Tsotsos, 2020, p. 57). Beyond the technical context, the technique has also been described as philosophically and psychologically essential for artificial intelligence: Dehaene et al. (2017, p. 492) argue that the underlying principle is necessary for a machine to "behave as though it were conscious", and both Samsonovich and Nadel (2005, p. 672–673) and Castelfranchi and Falcone (2019, p. 1) view it as a fundamental principle of "the self" that artificial selves must replicate. Section 6.1 of this paper addresses further purposes as stated in the relevant reports.

---

[1]Here, we set aside the metaphysical debate of whether machines may be capable of human-like thought. CCA research generally accepts this as a useful metaphor (Kotseruba and Tsotsos 2020, p. 21). Section 2.1 adopts a functional definition of cognition by Sanz et al. (2007), which elegantly avoids this issue. This perspective necessarily excludes dualist views that deny machine consciousness, but enables clear eligibility criteria for our review. For broader discussion see Berkeley and Rice (2012) and Dehaene et al. (2017).

[2]Earlier systems already meet our later definition of metacognitive experiences — e.g., PRODIGY (Carbonell et al. 1991) — but, to our knowledge McCarthy (1995, p. 11) was the first to state explicitly that"keeping a journal of physical and intellectual events so it [the robot] can refer to its past beliefs, observations and actions" is a prerequisite for introspective consciousness (in robots).

Two major issues with previous work on this topic are evident (for which we will present further evidence in this paper): First, many of these CMAs draw inspiration directly from human cognition instead of incorporating approaches developed within earlier CMAs, repeatedly reinventing the wheel. Secondly, the traces themselves are not well studied; Gordon et al. (2011, p. 296) attribute this to research focusing on algorithms rather than representation, resulting in models that are "modular, narrowly scoped, and specifically tied to particular agent architectures". The various approaches differ not only in their purpose and applications of metacognitive control but also in the data traced, abstraction level, processing algorithms, terminology, and, of course, the particularities of the respective CMAs.

Note that such problems also apply to (computational) metacognition in general, as has been noted previously, for example, by Cox and Raja (2011a, p. 3):

> The term [metacognition] is an overloaded one, and no consensus exists as to its definition. [. . . ] Indeed, Ann Brown (1987) described research into metacognition as a 'many-headed monster of obscure parentage.' Many of the technical terms used in research on metareasoning and related areas are quite confusing. Often, authors use different terms for the same concept (e.g., introspection and reflection), and sometimes the same terms are used in different ways (e.g., metareasoning has been cast as both process and object).

Efforts to slay Brown's metaphorical monster and to impose order on this landscape include a manifesto (Cox and Raja 2011b), a metamodel[3] (Caro Piñeres et al. 2014), an extension (Kralik et al. 2018) of Laird et al.'s (2017) *Common Model of Cognition*, as well as a review (Kotseruba and Tsotsos 2020), overviews (Anderson and Oates 2007; Cox 2005a,b; Cox et al. 2021), and comparative analyses of CCAs (Caro Piñeres and Builes 2013; Chong et al. 2009; Thórisson and Helgason 2012), that touch on metacognitive concepts. However, these works tend to be unsystematic or so broadly focused that findings stay fairly general. Specific techniques — including the one studied in this work — are, at best, only superficially covered.

In this paper, we approach metacognition from a more directed angle, systematically investigating how CMAs represent their metacognitive experiences, maintain them in episodic memory, and later analyze them to improve overall system performance. We use this technique as an organizing principle to guide our selection of literature and to delimit our analysis. The research goal is best summarized in the (oddly elegant and in itself somewhat "meta") research question: *How do models of cognition model their cognition?* This question is central to cognitive architecture research; Sanz et al. (2007, p. 940) postulate that the quality of the models employed by cognitive architectures directly limits the architectures' performance.

Although our focus is on a specific aspect of metacognition in CMAs — namely, how they retain and utilize metacognitive experiences — this capability is critical for many contemporary approaches. Whereas CMAs that lack an episodic memory discard monitoring data immediately after real-time processing, our work zeroes in on those that preserve their metacognitive experiences. This deliberate restriction is necessary to provide a detailed study rather than another superficial overview. To our knowledge, this work is the first with a sufficiently targeted scope to systematically analyze a key dimension of CMA metacognition in depth, further highlighting how fragmented the CMA research landscape is. By addressing the research question posed, we hope not only to gain insight into how CMAs function but also to reflect on the research community's understanding of human cognition. After all, a CMA's monitoring of cognition and the resulting mental simulations are ultimately inspired by — or even modelled after — our own thought processes. Indeed, increased efforts in CMA research are imperative to complete our picture of the mind (Langley 2017, p. 4875).

[3] As standardized by the *Object Management Group* (OMG); see https://www.omg.org/spec/MOF

The remainder of this paper proceeds as follows. After Section 2 details the applied methodology, we present and interpret our results Sections 3 to 6, each with a different focus. Section 7 discusses key themes and limitations emerging from this analysis, alongside implications for future research. Inspired by Kotseruba and Tsotsos' (2020) overview, we visualize much of the data collected in this review and offer interactive versions of these figures on a companion website[4] that allow, e.g., to filter the presented data by the CMAs. Furthermore, the raw data collected in this review are available in a GitHub repository[5].

# 2 Methods

This chapter includes (2.1) a concretised scope, (2.2) the eligibility criteria for papers to be included, (2.3) the search for relevant papers, and (2.4) the screening process, (2.5) a description of how they were analysed, and, finally, (2.6) how we present the synthesis.

## 2.1 Scope

As outlined in the Introduction, we are interested in CMAs that maintain an episodic memory of metacognitive experiences in order to improve their overall performance. Here, we specify the relevant terminology, clarify the scope, and differentiate both from related topics.

### 2.1.1 What is Metacognition?

Distinguishing cognition from metacognition is not always straightforward (Herrmann 2023, p. 12). Here, we clarify our earlier intuition of what metacognition is — and by extension, about CMAs — that we gave in the Introduction. Conventionally, the prefix "meta" signifies a concept applied to itself; that is, "meta-X" means "X about X" (Lehrer 1995, p. 136). In the case of metacognition, we are literally dealing with cognition about cognition; therefore, we must first characterize cognition. Among the numerous definitions available, we adopt the following because it provides a precise and functional framework that aligns with the objectives of our research (as will soon become evident):

**Definition 1** [*Model-Based Cognition*; after Sanz et al. (2007, p. 939)]**.** A system is *cognitive* if it exploits models of other systems in interacting with them.

It is important to note that the above definition technically narrows the broad understanding of CCAs identified in Kotseruba and Tsotsos' survey (2020, p. 22). Because human cognition meets Definition 1, it stands to reason that CCAs as software that mimics human cognition, including mental representations, should also satisfy Definition 1 transitively.

[4] https://dml.uni-bremen.de/ease/review
[5] https://github.com/mrnolte/cma_review_data

Building upon the principle of Model-based Cognition, Sanz et al. (2007, pp. 943–944) implicitly capture the concept of metacognition as well. We make this connection explicit.

**Definition 2** [*Model-Based Metacognition*; adapted from Sanz et al. (2007, pp. 943–944)]**.** A cognitive system is *metacognitive* if it exploits self-models in interacting with itself.[6]

Nelson and Narens (1990, p. 126) describe a meta-level *mental simulation* as a human's representational model of their object-level cognition. This concepts aligns with the "self-model" required in Definition 2, while metacognitive control equates to self-interaction. In a CMA, its self-model and self-regulation serve the same roles. The term "self-model" here covers both symbolic and sub-symbolic forms — for example, a knowledge graph of software components or system behaviour statistics. The principle of Model-Based Metacognition posits that underlying self-models are the defining feature of metacognition, underscoring our research question ("How do models of cognition model their cognition?"). As the principle's contraposition suggests, metacognition cannot exist without such self-models.

### 2.1.2 What are Metacognitive Experiences?

Many authors have explored facets of metacognition beyond our Definition 2. One of the earliest and most influential distinctions comes from Flavell (1979), who classifies three types of data linked with metacognition [paraphrased from our previous work (Nolte et al. 2023)]:

*Metacognitive knowledge*[7] refers to factual information about one's own cognition; for example, knowing that you understand how to solve for $x$ in an equation.

*Metacognitive strategies* are subconscious behaviours for manipulating one's own mind, such as the learned, automatic habit of creating mnemonics to aid memory.

*Metacognitive experiences* are "any conscious cognitive or affective experiences that accompany and pertain to any intellectual enterprise" (Flavell 1979, p. 906).

Some literature on human cognition equates metacognitive experiences with emotions such as feelings of confidence [e.g., (Schwarz 2010)]. For the purposes of this review, we understand Flavell's definition more broadly and reapply the principle underlying Definition 2:

**Definition 3.** *Metacognitive experiences* are internal models of a (meta)cognitive system's own internal processes or states, which the system can deliberately access and inspect.

Let us forestall the criticism that undergoing metacognitive experiences should not be ascribed to artificial agents. Block (2002) distinguishes between the concept of consciousness in the experiential, phenomenal sense (e.g., the "totality" of what it is like to taste something spicy) and the concept of consciousness as an accessible representation of the content of the experience (e.g., the logical statement "the food is spicy" for the subsequent planning to add yoghurt). Henceforth, when speaking of metacognitive experiences in the context of CMAs, we always mean the representational sense specified by Definition 3 unless we state otherwise[8]. This also limits Flavell's (1979) original account of metacognitive experiences, which encompasses both phenomenal and representational aspects, to the latter. One might further argue that experience is then the wrong term and that *meta-representation* would be more suitable. Yet not every meta-representation is metacognitive. For example, if an apple icon represents an apple, then perceiving that icon involves a meta-representation of an apple. Furthermore, the metaphorical use of the term *(metacognitive) experience* coincides with the usual appropriation from human cognition in CCA research, as when a database is described as storing the "the personal experience of the system" (Kotseruba and Tsotsos 2020, p. 45).

[6] By "(it)self", we mean a mind — human or computational (i.e., a CCA) — as a system separate from its possible embodiment. This restriction excludes body models, which pertain to object-level cognition and therefore fall outside Definition 2.

[7] Not to be confused with (object-level) metaknowledge; cf. Cox (2011, pp. 134–136).

Hence, metacognitive experiences may be viewed as the conscious versions of the self-models required by Definition 2. This perspective aligns with the psychological consensus that metacognition necessitates "representations of one's own first-order mental states" (Fletcher and Carruthers 2012, p. 1368). The processes described in Definition 3 are flexible in their degree of abstraction, encompassing an intentionally broad spectrum of phenomena ranging from high-level cognitive functions such as planning to more lower-level experiences, including individual qualia or percepts. Importantly, not all metadata managed by a meta-level qualifies as a metacognitive experiences or trace thereof: Soar and ACT-R, for instance, track activation metrics that are not consciously accessible to the agent (Laird 2021, p. 4).

To keep this review feasible, we only consider a CMA's memories of their metacognitive experiences. Consider the following illustration: Alice plans to see a friend on Saturday, and because the forecast predicts sunshine, they agree to meet in a park. As the date nears, the forecast shifts to rain. Recognizing this, Alice proposes moving their meeting to an indoor venue. In this scenario, Alice retraces where she initially used the now-updated weather information and reevaluates the impacted decision — without having to update unaffected choices. To apply this example to CCAs, imagine a household robot that reschedules washing and hang-drying laundry for similar reasons. To adjust its schedule appropriately, the robot must retain traces of its planning procedures in episodic memory. There may be additional dependent decisions; for instance, the robot might need to reschedule other tasks to fill the newly available time slot. The agent's memories of their "train of thought" in both examples clearly satisfy Flavell's (1979, p. 906) original definition and, of course, our Definition 3.

To be precise, we require that the kept metacognitive experiences have an episodic character that would permit some form of *autonoetic consciousness* or *mental time-travel* (Michaelian 2016; Sant'Anna et al. 2024), either retrospective (memories) or prospective (imaginations). For instance, simply associating arising thoughts with feelings of confidence without any temporal context is insufficient. Even a comprehensive model of the object-level system state limited to the present, such as that of TOMASys (Hernández Corbato 2013), is not relevant to this review. The kind of autobiographic experiences we require are typically associated with what Tulving (1972, 1985) coined as *episodic memory*, a concept on which CCA research has yet to reach a consensus (Laird et al. 2017, p. 22). Indeed, episodic memory "has been largely ignored" by CCAs (Kelley et al. 2018, p. 718) and "remains relatively neglected in computational models of cognition" (Kotseruba and Tsotsos 2020, p. 70). Although exceptions exist, such as the CMAs Soar (Laird 2019) and CRAM (Beetz et al. 2010), we therefore do not require the retention of traces to be explicitly *framed as* episodic memory. For example, the developers of Meta-AQUA reject that the CMA features an episodic memory (Cox et al. 2011, p. 78), a claim we accept for object-level experiences. However, Meta-AQUA generates reasoning traces that meet our definition of metacognitive experiences and temporally link events, thereby justifying inclusion in this review. In the remainder of this paper, we also synonymize the episodic metacognitive experiences described in this paragraph as "traces" and "memories" of metacognitive experiences.

[8] A similar method was proposed by Castelfranchi and Falcone (2019, p. 1–2).

We noted in the Introduction that there is no terminological consensus among CCA regarding the concepts discussed here — a point for which we will present evidence in Section 4.2. Our account of the underlying theory derives the vocabulary from psychology, specifically the term "metacognitive experiences". While the conceptualization in Definition 3 informs our overall approach, we deliberately do not carry this specific terminology into the practical implementation of our methodology, e.g., into the search query. Rather, consider the terminology we use in this section — and throughout the remainder of this paper — as a medium to ensure a clear, uniform narrative, independent of the heterogeneous terms found across the literature. The latter is instead explicitly synthesized in Section 4.2.

### 2.1.3 Related Topics

We laid out some work related to this review in the Introduction. This section briefly explores approaches and concepts from adjacent fields that closely resemble our technique of interest, noting any significant differences. For additional insights into other systems that employ traces analogously to how CMAs use them for metacognition, such as path planners, see Cox (2005a, p. 9–10; 2005b, p. 126). The diversity of these approaches underscores the interdisciplinary nature of our review's topic. Nonetheless, to maintain a focused scope, we include only systems in the review that are explicitly presented as CCAs or CMAs.

*Automated log analysis* transforms software log messages — typically unstructured text — to machine-readable events, e.g., via machine learning (Gholamian and Ward 2021). This process resembles metacognitive monitoring: the software instance serves as the object-level, and automated log analysis as the meta-level. In particular, it parallels the diagnoses phase explained in Section 6.3 and shown in Figure 14 on page 39. Similarly, *software tracing tools* record event chains and telemetry within complex software systems, facilitating later analysis by developers, administrators, or customer support personnel (Janes et al. 2023). These analyses are generally external to the system. Thus, both automated log analysis and software tracing tools typically lack the aspect of self-regulation inherent in metacognitive control. So-called *self-healing systems* (Schneider et al. 2015) are an important exception to this.

*Multi-agent system frameworks* orchestrate the collaboration of various software agents to solve complex tasks, often monitoring message exchanges and internal events among agents (Sturm and Shehory 2014). Mature CMAs such as Soar (Laird 2012) fit this category. Notably, both tools for software tracing and multi-agent systems, within each category and akin to the phenomenon we observe with CMAs, display conceptual homogeneity yet employ heterogeneous terminology (Janes et al. 2023; Sturm and Shehory 2014).

Some *semantic logical reasoners* and *planners*, such as Krulwich's (1991) CASTLE system, produce reasoning traces and justification structures that can resemble traces of metacognitive experiences (Cox 2005b, p. 126). *Case-Based Reasoning* (CBR), as well as Carbonell's (1985) *derivational analogy*, even leverage previous solutions to solve novel problems (Bartsch-Spörl et al. 1999). When these systems store and recall not just solutions but also representations of the processes that generated them, they may satisfy our Definition 3 of metacognitive experiences. Some surveyed CMAs [PRODIGY (Carbonell et al. 1991); Metacat (Marshall 2006); ACNF (Crowder et al. 2014); CARINA (Caro Piñeres et al. 2019)] use CBR in this manner. However, many CBR systems are not framed as CCAs or CMAs and remain excluded from our study. Similarly, some systems like Stroulia's (1994) AUTOGNOSTIC are not framed as CCAs or CMAs but use still a form of *error-driven learning* by tracing employed problem-solving algorithms. In cases of failure, these traces facilitate blame assignment so that the system may initiate countermeasures, e.g., learning processes. Indeed, one CMA in this review — Meta-AQUA (Ram and Cox 1995) — adopts a parallel approach.

Our Definition 3 of metacognitive experiences resembles the influential consciousness concept of *C2 qualia* by Dehaene et al. (2017) and Peters (2022, p. 7). The former define C2 "by the ability to reflexively represent oneself" (Dehaene et al. 2017, p. 400), i.e., as the self-monitoring aspect of consciousness, juxtaposed to its counterpart C1, which they describe as the global availability of information across all cognitive functions. This distinction mirrors Nelson and Narens's (1990) object-level (C1) and meta-level (C2) classification.

Van Ments and Treur (2021) provide a brief overview of *mental models*, originally introduced in the 1940s, to represent or imagine external world states. These models have since been extended to describe mental processes, such as reasoning states and their transitions.

We here skip *Chain-of-Thought* (CoT) techniques that have seen increased usage in LLMs over recent years and instead discuss them in the Outlook section.

## 2.2 Eligibility Criteria

Having established the scope, we now introduce our eligibility criteria for this review. A report is considered eligible only if it satisfies all of the following conditions:

---

*Relevancy:* The report describes a concrete technique for CMAs to monitor, model, and inspect episodic metacognitive experiences at runtime to improve the CMAs' performance.

*Quality:* The report has been published in a scientific journal or book, at a scientific conference or workshop, or as a dissertation.

*Language:* The report is primarily written in English.

---

The Relevancy Criterion ensures that reports fall within the scope defined in Section 2.1. The Quality Criterion aims to uphold scientific rigour: we expect reports that pass this filter to have undergone peer review or editorial scrutiny, though verifying this for each report is infeasible. If a paper appears notably poor quality, we will investigate whether it was published in a predatory journal or at a disreputable venue; if so, it will be excluded from our review. Lastly, the Language Criterion ensures accessibility to a general audience[9]. Note that the criteria do not differentiate between primary and secondary literature. We will discuss overviews, meta-analyses, and systematic reviews separately in Section 3.1.

## 2.3 Paper Search

We began by conducting an exploratory literature search, guided by key descriptions of our technique of interest (Cox, 2005b, p. 123–126; Gordon et al., 2011, p. 296; Kotseruba

---

[9] Following Dawidpuljak's (2021) guidelines, we treat language as an eligibility criterion rather than a search query filter.

```
(metacognition OR metareasoning) AND (monitor* OR trace*) AND
("cognitive_agent" OR "cognitive_architecture" OR
 "metacognitive_architecture")
```

**Listing 1**The query $q$ defined to search for relevant records

and Tsotsos, 2020, p. 57)[10]. We quickly encountered the problem that the technique lacked a universally accepted name, complicating the identification of relevant reports. While "metareasoning" is prevalent in logic literature, CMAs papers often use broader terms such as "metacognitive monitoring and control", which can encompass unrelated techniques (e.g., allocating tasks among internal agents based on varying capabilities). Other authors use different terminology; for instance, Castelfranchi and Falcone (2019, p. 3) describe agents that "meta-examine" their reasoning. In our initial literature subset, the data structures generated during the process were sometimes referred to as "traces" — or, more precisely, "reasoning traces". Notably, labels such as "meta-representation" and "episode" only became prevalent later in our screening process. Consequently, they were not included in our search query.

Based on these findings, we iteratively developed our final search strategy, $q$, which is shown in Listing 1. We tested and refined it across several academic databases (the ones we later screened; see Figure 1). The first two disjunctive clauses in $q$ aim to focus on the technique of interest despite the vague and heterogeneous terminology: "monitoring" is the standard term used to describe the flow of information from the object-level to the meta-level, while "trace" extends this clause with a more technical term that is mainly used in computer science. Given this inclusive approach, we refrain from further iteration with the more specialized terms identified later (Section 4.2). The third disjunctive clause excludes results outside computational cognitive architectures — such as works in philosophy or animal psychology. Many modern academic databases support query expansion to match alternative spellings and inflections, especially via wildcards using the $*$-operator (Gusenbauer and Haddaway 2020). For instance, Google Scholar also matched terms including "meta reasoning", "meta-cognitive", "monitors", and "tracing".

To choose appropriate databases for executing $q$, we referred to Gusenbauer and Haddaway' (2020) evaluation of 28 academic search engines for systematic reviews. Based on carefully selected criteria, they classified the databases into *principal* and *supplementary*, recommending that the latter be used only as secondary resources. Since their study's publication, Microsoft Academic and WorldWideScience.org have been discontinued. We then discarded eight databases — Clinical-Trials.gov, Cochrane Library, EBSCOhost, ERIC, OVID, ProQuest, PubMed, TRID, and Virtual Health Library — that reportedly do not cover computer science. We also excluded CiteSeerX because it lacks support for Boolean operators[11].

Figure 1 presents the number of results returned by $q$ for each of the six principal and ten supplementary databases remaining, using the standard search fields without limits. We omitted wildcards from $q$ when searching in Science Direct due to its lack of support. We could not replace wildcards with an explicit disjunction (e.g., traces OR traced OR trace OR tracing) because Science Direct limits the number of Boolean operators per query. In three

[10]The relevant reports found in this process were also included in our review (see Section 2.4).

[11]CiteSeerX returned 9 639 051 records for $q$. A test query for **OR AND** resulted in 9 638 357 records — less than $\leq 0.01\%$ fewer. Because screening such a volume is impractical, we excluded this database.

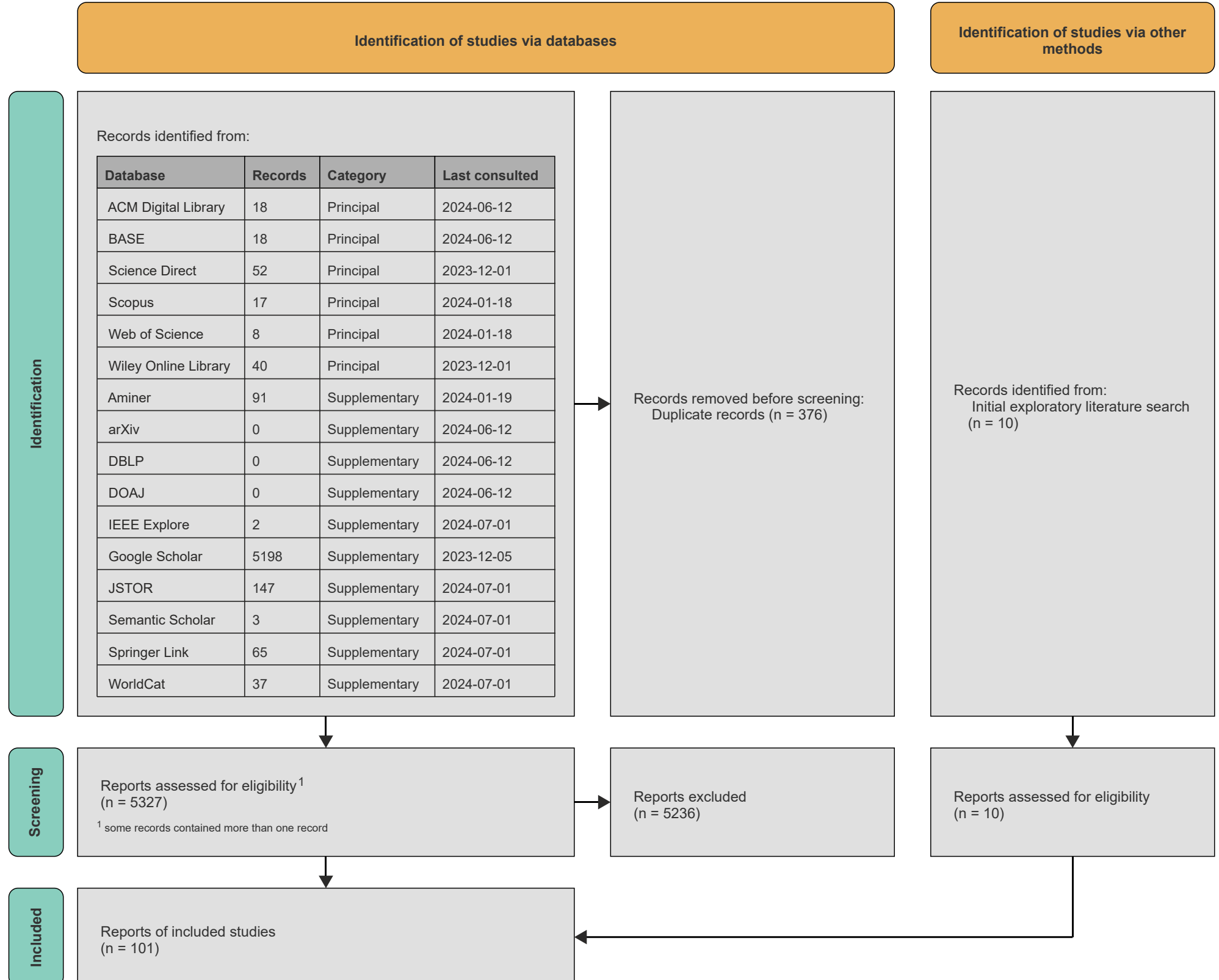

| Database | Records | Category | Last consulted |
|---|---|---|---|
| ACM Digital Library | 18 | Principal | 2024-06-12 |
| BASE | 18 | Principal | 2024-06-12 |
| Science Direct | 52 | Principal | 2023-12-01 |
| Scopus | 17 | Principal | 2024-01-18 |
| Web of Science | 8 | Principal | 2024-01-18 |
| Wiley Online Library | 40 | Principal | 2023-12-01 |
| Aminer | 91 | Supplementary | 2024-01-19 |
| arXiv | 0 | Supplementary | 2024-06-12 |
| DBLP | 0 | Supplementary | 2024-06-12 |
| DOAJ | 0 | Supplementary | 2024-06-12 |
| IEEE Explore | 2 | Supplementary | 2024-07-01 |
| Google Scholar | 5198 | Supplementary | 2023-12-05 |
| JSTOR | 147 | Supplementary | 2024-07-01 |
| Semantic Scholar | 3 | Supplementary | 2024-07-01 |
| Springer Link | 65 | Supplementary | 2024-07-01 |
| WorldCat | 37 | Supplementary | 2024-07-01 |

**Fig. 1** PRISMA 2020 flowchart (Page et al. 2021) adapted to our procedure. An interactive version on the companion website links directly to every database query

databases, we used discipline-specific filters, selecting "Computer Science" and "Engineering" for Science Direct and Springer Link, and "Computer Science" and "Electrical & Electronics Engineering" for Wiley Online Library. In Google Scholar, we excluded cited references.

Google Scholar was particularly sensitive to synonyms — such as "display" for "monitor". This automatic query expansion was unavoidable and led to many irrelevant results. Because Google Scholar also generally covers more papers than other engines (Gusenbauer and Haddaway 2020), it returned a substantially larger number of results. In contrast, Aminer, DBLP, and Scopus lack full-text search options and yielded relatively low result counts.

To capture as many relevant reports as possible, we decided to screen the search results of all 16 databases shown in Figure 1, thus covering a broad mix of both principal and supplementary databases as classified by Gusenbauer and Haddaway (2020). This strategy produced 5,689 search records across all databases. A cross-database script excluded 369 duplicates, leaving us with a total of 5,320 records.

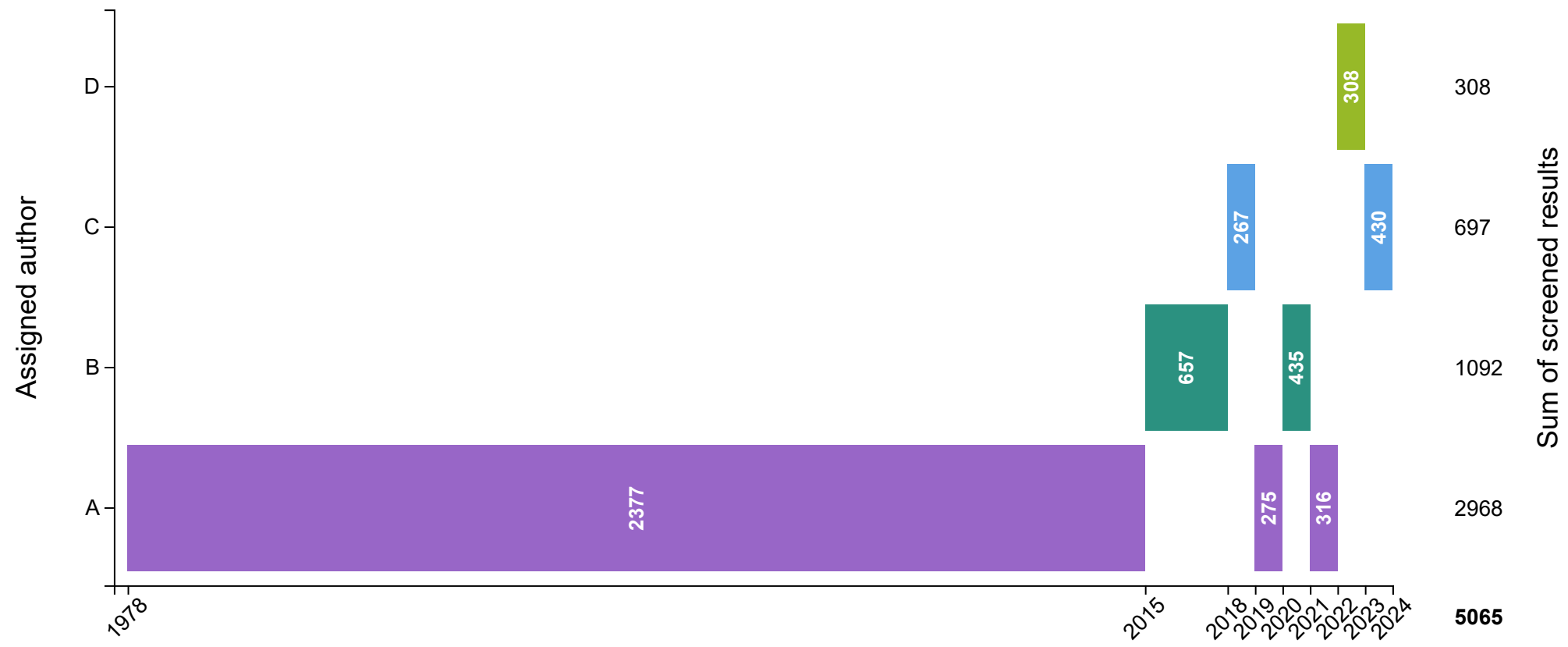

**Fig. 2** The timespans of Google Scholar search records screened by authors (A–D), labelled with the number of search records

## 2.4 Screening Process

Five authors (screeners A–E) carried out the screening manually and independently. Each record was assigned to a single screener as follows:

- Screener A handled ACM Digital Library, BASE, Scopus, Web of Science, Wiley Online Library, Aminer, IEEE Explore, JSTOR, Semantic Scholar, Springer Link, and WorldCat.
- Screeners A-D divided Google Scholar results by publication-year ranges[12] as shown in Figure 2.
- Screener E handled Science Direct.

The screening phase lasted roughly eight months, during which we re-ran $q$ regularly against all databases to capture new entries added to indexing databases such as Google Scholar.

An initial inspection revealed that many results were clearly irrelevant due to being from educational science, philosophy, human psychology, or medicine. Screeners were therefore allowed to directly classify a record as irrelevant if they were confident from the title and abstract that it did not concern CCAs. Records on CCAs and metacognition were harder to classify: Our relevance criterion — although aimed at one specific technique — is phrased in a general way. Data gathered by many CCA was non-episodic or concerned only external events. Abstracts and titles rarely sufficed to identify this, and screeners then had to check via a time-consuming read-through of the full text.

We initially classified 107 records as relevant. Of these, 97 arose from the systematic screening, and 10 from the initial paper search (Section 2.3). Note that 4 of the 107 records point towards proceedings or collections that contained at least one other relevant report.

During our thorough read-through, we noticed that three of the reports initially deemed relevant were of remarkably poor quality. We investigated these papers and found that the publishing journal appeared both on Jeffrey Beall's (admittedly outdated) list of predatory journals[13] and the *Directory of Open Access Journals*'s (DOAJ's) list of fraudulent titles (DOAJ 2014). We therefore discarded them in accordance with our Quality Criterion.

[12]By examining the Google Scholar results year-by-year, we circumvented the 1 000-hit limit.

The team marked 17 further search records as "unsure". After group discussion, all of them were discarded as ineligible. Additionally, one record pointed to a preprint (Kotseruba and Tsotsos 2018), which met the relevance and language criteria but initially failed the Quality Criterion due to its preprint status. Since it was subsequently published in a peer-reviewed journal, we included the later version (Kotseruba and Tsotsos 2020) in our analysis.

The final corpus comprises 101 relevant reports (excluding the four collections; see Figure 1). A complete list — some items are absent from the main bibliography — is available on the companion website and in the data repository.

## 2.5 Data Extraction

To address the general research question posed in the Introduction systematically, we formulated 20 subsidiary questions for data extraction. We group them into three categories (A) *Model Context*, (B) *Data Content and Structure*, and (C) *Objectives and Evaluation*:

(A) **Model Context**

- (A1) What Computational Cognitive Architecture (CCA) is used?
- (A2) When was the report published?
- (A3) How is the technique of interest referred to (e.g., tracing, meta-examination, or vaguely monitoring and control)?
- (A4) Is the study theoretical or implemented?
- (A5) Which non-computational models of human cognition from cognitive psychology have inspired the study, if any?
- (A6) Have other CCAs, e.g., predecessors or competitors, inspired the study?
- (A7) If so, what are the main innovations compared to the inspiring studies?

(B) **Data Content and Structure**

- (B1) Is the traced data symbolic, sub-symbolic, or hybrid?
- (B2) What information does the traced data cover, and what does it explicitly not cover?
- (B3) What is the level of detail traced (e.g., higher-level cognitive functions such as Reasoning vs. single algorithm steps)? Are there examples?
- (B4) Is there a formal specification of the traced data, such as mathematical or ontological definitions, and if so, what?
- (B5) Is the covered data system specific (e.g., are processes traced by method names), or is it general (e.g., inspired by folklore psychology)? Are there examples?
- (B6) How is the data structured (e.g., is it continuous information or just specific episodes? Are those linked to each other?)?
- (B7) How is the data structure implemented and stored in memory (e.g., via databases, linked lists, or knowledge graphs)?

(C) **Objectives and Evaluation**

- (C1) Is the work framed towards a specific purpose (e.g., explainability, consciousness, or autonomy)?
- (C2) Which particular issues are to be improved?
- (C3) Are there any worked examples for illustration?

[13] Although Beall removed his list from the internet in 2017, mirror sites persist. Critics such as Teixeira da Silva (2017) argue that the lack of transparent inclusion criteria make the list discriminatory.

| Screener | A | B | C | D | E |
|---|---|---|---|---|---|
| **Examined reports** | 77 | 16 | 14 | 12 | 0 |

**Table 1** The number of examined reports per screener

(C4) Which algorithms process the traced data (e.g., machine learning or trace inspection)?
(C5) Are there any test scenarios outlined?
(C6) Are the techniques evaluated? If so, how successful are the applied methods in solving the issues or improving the CCA's performance?

We derived the questions above from our initial literature search, although they bear some similarity to those defined by Zilberstein (2011, p. 32–33). Zilberstein, however, focuses on contrasting a CCA's performance gains from metacognitive strategies with the attendant computational overhead, which is not our concern here. Instead, we aim to provide a structured overview of the fragmented research on memories of metacognitive experiences in CMAs. Each item serves a distinct purposes, which we outline briefly in Sections 4 to 6. Where the synthesis warrants it, we may extract additional data not explicitly covered by these questions. Figure 9 exemplifies this: the first row covers Item (A4), whereas the second row records whether or not the corresponding code is publicly available.

We started by sorting the relevant reports by architecture. If present, we grouped relevant reports by monikers for the studied CMA(s) as designated by their authors. We further grouped reports on closely related CMAs into "families" for comprehensibility; see Section 4.5 for details. Each family is assigned a consistent label of our choosing — for example, the *BICA family* includes, among others, *GMU BICA* (Samsonovich et al. 2009; Kalish et al. 2010) and *eBICA* (Samsonovich 2013a). For the minority of CMAs that are unnamed and stand-alone, we coin an acronym composed of the author's initials plus the publication years. Thus, the CMA presented by Brody et al. (2013) becomes *BCP13*. We also merge papers that appear to describe the same unnamed CMA, based on conceptual homogeneity, evolving ideas, and overlapping authorship or references; in such cases the earliest relevant report determines the label.

Screeners A–E were then assigned different CMAs to read the associated grouped reports independently and extract answers to questions (A1)–(C6). Table 1 shows the resulting report distribution. As five architectures were examined by two screeners, the sum of the assignments exceeds the total count of relevant reports. All extracted data items were recorded in a shared Google Sheets[14] spreadsheet; a copy of the raw file is available in the data repository.

## 2.6 Synthesis, Reporting, and Presentation

To present a coherent picture of the technique of interest, we organize the synthesis by questions (A1)–(C6) instead of by CMA. This choice risks repeating architecture-specific concepts and thereby confusing readers, given how many architectures we examined. To address this, we refer to the section that offers the most details rather than repeating ourselves wherever possible. An index of acronyms and terms of interest on page 52 — including every examined CMA's name, helps readers locate all occurrences of a given framework.

Wherever possible, we follow Kotseruba and Tsotsos (2020), who make their data accessible through interactive data visualizations. Our figure captions indicate when a corresponding interactive version is available on the companion website. Those visualizations, built with the *D3.js* library[15], allow, for instance, to filter the presented data by CMA.

[14] https://docs.google.com/spreadsheets/

Due to the qualitative nature of our data, a meta-analysis is infeasible. We primarily employ a structured narrative (thematic) synthesis approach to explore and summarize patterns in how different CMAs handle metacognitive experiences and to showcase concrete examples from each architecture. Where useful, we support these descriptions by grouping the CMAs into categories and comparing their frequencies. Should the data warrant more rigorous methodology, we may apply techniques such as semi-automatic clustering, genealogization, or timelining. Unless stated otherwise, narrative synthesis remains the default; any departure from the baseline method is noted in the respective sections.

The *2020 Preferred Reporting Items for Systematic reviews and Meta-Analyses* (PRISMA 2020) were developed for confirmatory reviews in the health sciences. As such, several checklist items presume data suitable for meta-analysis. Since our exploratory research question centres on qualitative evidence typical of scoping reviews, adjustments are necessary. For instance, we cannot perform a classical *risk of bias assessment* (PRISMA 2020 Items 11, 14, 18, 20a, and 21) and *certainty assessment* (PRISMA 2020 Item 15 and 22): problems such as lack of study randomization cannot occur when formal evaluations are absent [a gap explored under question (C6)], and the heterogeneous, engineering-oriented methods of the relevant reports preclude application of a standard risk-of-bias tool designed for clinical trials.

# 3 Preliminary Results and Overviews

## 3.1 Secondary Literature

Besides the CMA-specific reports, we identified four broader relevant publications.

Castelfranchi and Falcone (2019) provide a philosophical account of a robot's "self", linking the technique of interest to consciousness and intentionality, yet describing no concrete CCA.

Anderson and Oates's (2007) overview of metacognition in computer science discusses reasoning with meta-representations. They focus on logics designed for metareasoning — such as *Active Logic* [cf. (Anderson et al. 2008; Brody et al. 2013; Goldberg 2022)] — and discuss the computational complexity and paradoxes that arise in self-referential logic.

Langley et al. (2009) give an overview of selected cognitive architectures, briefly noting how PRODIGY (Carbonell et al. 1991) employs traces, which we will examine in later sections.

In their exploratory overview, Kotseruba and Tsotsos (2020) investigate, among other aspects, how CCAs employ metacognition, observing that "some architectures support a temporal representation (trace) of the current and/or past solutions" (Kotseruba and Tsotsos 2020, p. 57). Although they discuss only the general purpose of such traces[16], their paragraph cites six CCAs; four of which are included in this systematic review (Metacat, MIDCA, FORR, and Soar). The other two [Companions and GLAIR] are excluded for separate reasons. Companions (Friedman et al. 2011) logically models alternative scientific theories (e.g., explanations of seasonal change), and compares their validity. Definition 3, however, requires a model of the system's *own* cognition, and Friedman et al. (2011, p. 121) concede that Companions "does not directly monitor the domain-level reasoning". GLAIR employs CBR but, to our understanding, without an episodic character as demanded by our

[15] https://d3js.org

[16] Recall that we mentioned the purposes identified by Kotseruba and Tsotsos (2020) in the Introduction.

Relevancy Criterion. In Shapiro and Bona's (2010) example, GLAIR adds shortcut rules to the knowledge base; for instance, after applying and expanding *transitive*$($*ancestor*$)$, GLAIR stores *ancestor*$(x, y) \wedge$ *ancestor*$(y, z) \implies$ *ancestor*$(x, y)$ as a single rule.

## 3.2 Data Availability

We classify the quality of the collected data in three categories:

*Detailed:* sufficient information to understand the approach thoroughly.
*Partial:* some information is available, but, to our judgement, not enough for replication.
*None:* no relevant information could be extracted.

Figure 3 lists the quality for each CMA and data item (A1)–(C6). Across all item-CMAs pairs, $(46.3\%)$ were rated Detailed and $21.2\%$ Partial. The remaining $32.5\%$ fell into the None category. Missing data were especially common ($\geq 50\%$ of CMAs) for level of detail (B3), formal specifications (B4), data specificity (B7), example scenarios (C3), test scenarios (C5), and evaluation (C6). Conversely, more than half of the CMAs provided Detailed information for the publication date (A2), terminology (A3), implementation status (A4), inspiring cognition models (A5), related architectures (A6), data type (B1), data content (B2), and purpose (C1).

Data availability also varies by architecture. Consider Item (C4) regarding algorithms: While the relevant reports explain, e.g., Meta-AQUA's algorithms step-by-step and even provide pseudo-code (Cox 1994), we found hardly any details for REM — only that it "would use [...] the trace of processing that led to the failure, to diagnose the causes of the error" (Goel and Jones 2011, p. 154). This contrast partly reflects an uneven literature base — in the previous example, 21 reports on Meta-AQUA versus one on REM — and partly differing emphasis: for instance, Meta-AQUA centres on trace processing, while BICA papers focus on trace structure and representation within a ToM-enabled system. Although BICA has the second-largest body of relevant reports (11, tied with MIDCA), we still found scant detail on how its traces are processed (see Section 6.3.4).

## 3.3 Research Community

To determine which CMAs were developed collaboratively by the research community, we analysed the partial co-authorship graph in Figure 4. A prominent component centres on Michael Thomas Cox, who appears as co-author on nearly 30% of all relevant reports. Collectively, this cluster accounts for over half of all relevant reports yet involves only 11 of the 35 CMAs (cf. Section 4.1): BCP13, CARINA, H-CogAff, KS02, K10, K12, K21, MCL, Meta-AQUA, MIDCA, and MISM. Some of these also share a close conceptual connection; for example, both MIDCA and MISM take inspiration from Meta-AQUA and MCL (cf. Section 4.5).

Outside the cluster, no author in our corpus has actively worked on more than one architecture; references to other CMAs appear solely for context. We take this as an indication that many researchers prefer independent designs or at least to integrate existing ideas into their own architectures rather than joining large collaborations. However, because the graph reflects only the relevant reports, additional publications could reveal further connections.

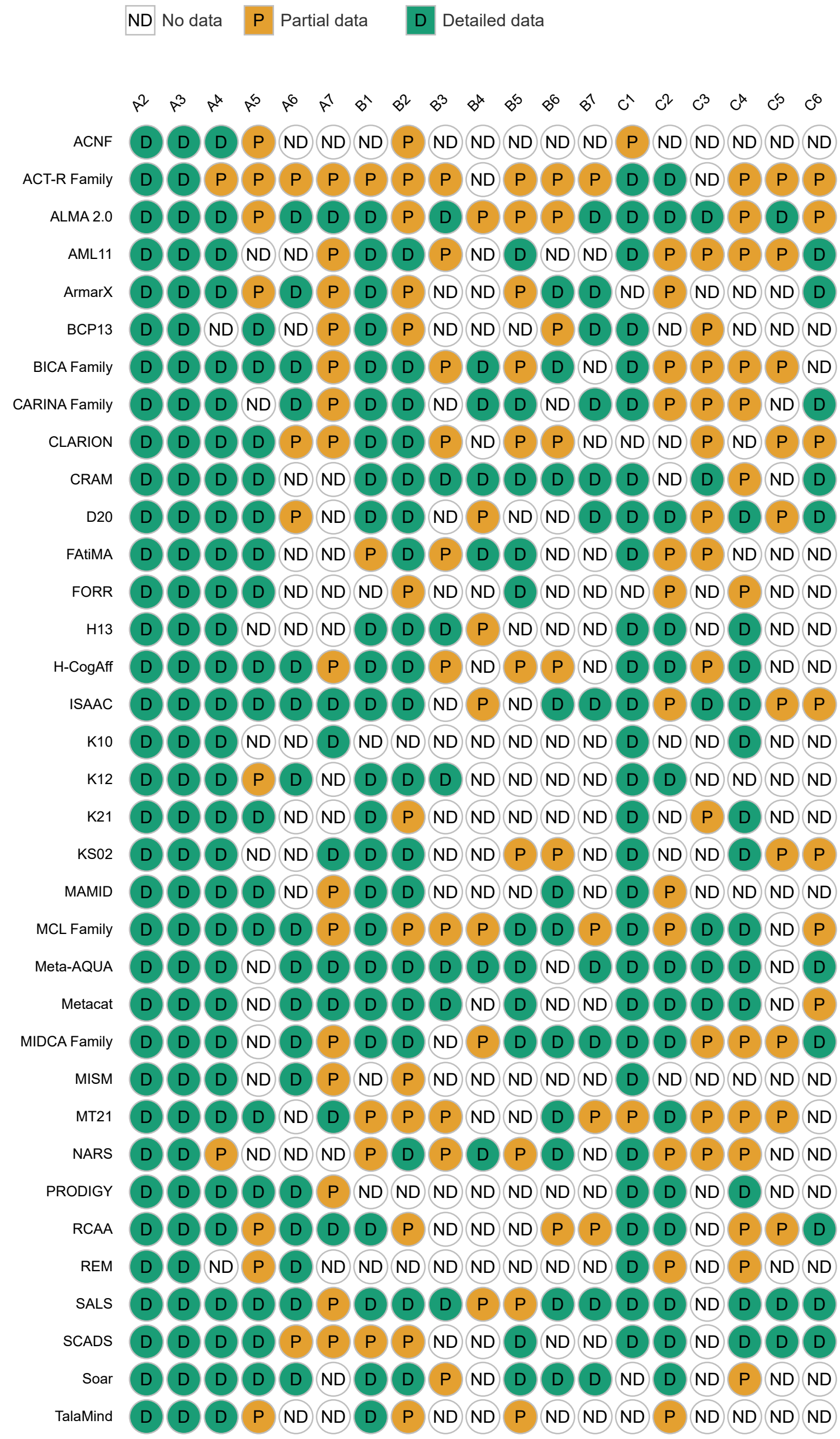

**Fig. 3** A grid graph depicting the quality of the extracted data for each reviewed CMA and data item. D = detailed data; P = partial data; ND = no data

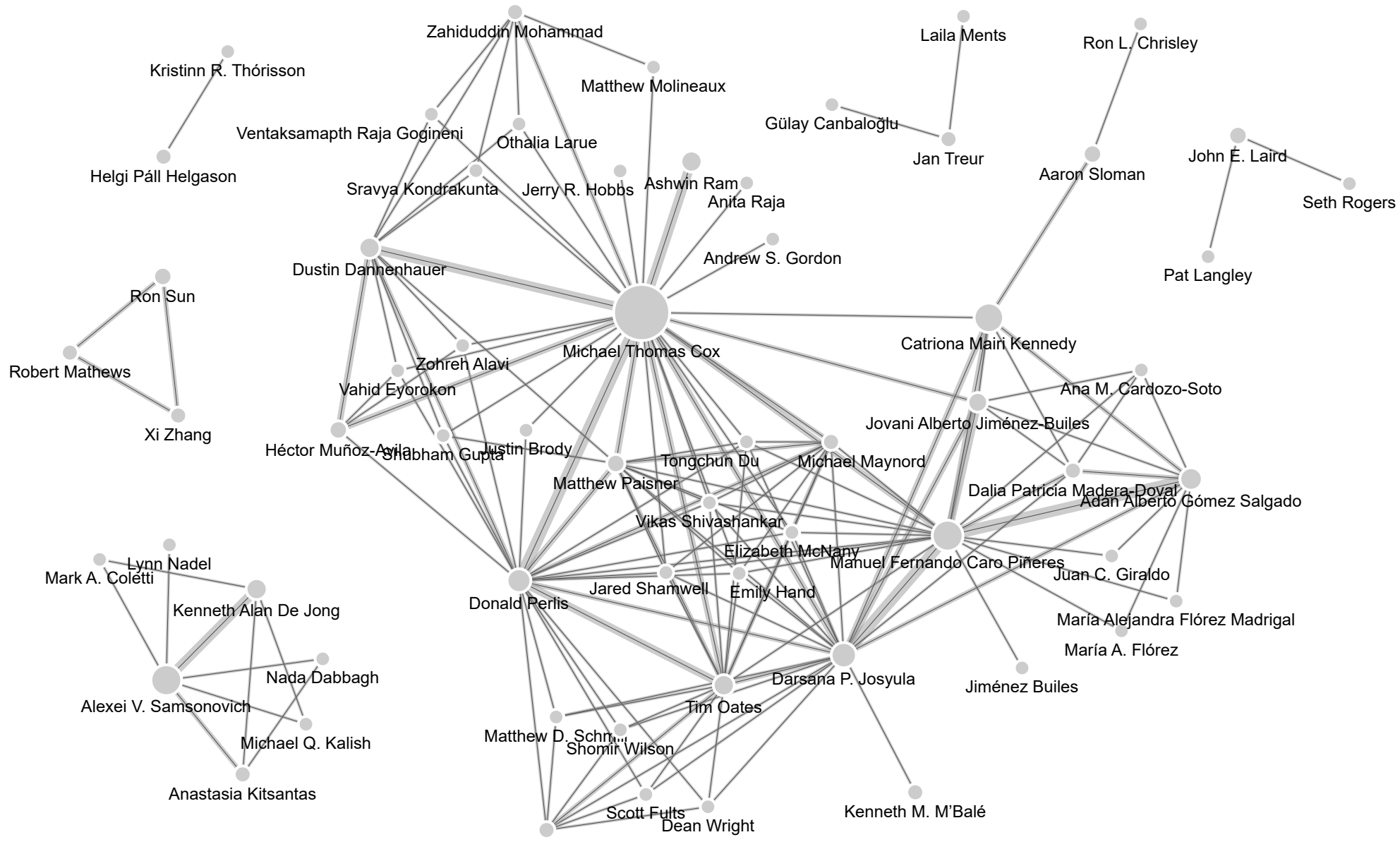

**Fig. 4** A partial co-author network of researchers who contributed at least one relevant report. Each node is labelled with the corresponding author's name, and node size scales with the number of relevant reports an author has produced. Edges connect co-authors, with thickness proportional to their shared publications. To maintain legibility in print, we consider only papers in which at least one author has more than one relevant report. The companion website offers an interactive version that involves all relevant reports

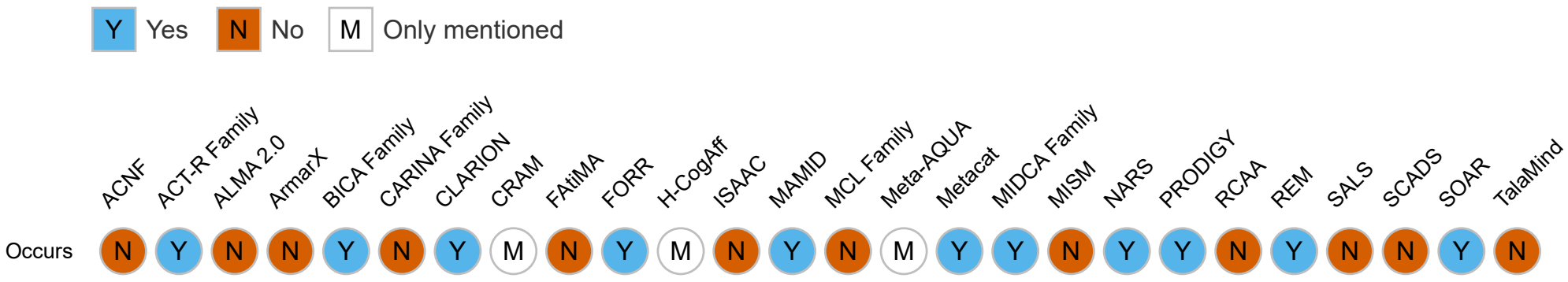

**Fig. 5** Whether the *named* CMAs that we consider in this review also occur in the overview by Kotseruba and Tsotsos (2020). Y = yes; N = no; M = mentioned, but explicitly not covered

Smaller clusters in Figure 4 correspond to Soar (John E. Laird), MT21 (around Jan Treur), BICA (around Alexei V. Samsonovich), CLARION (around Ron Sun), and H13 (Helgi Páll Helgason), and some of the secondary literature mentioned in Section 3.1.

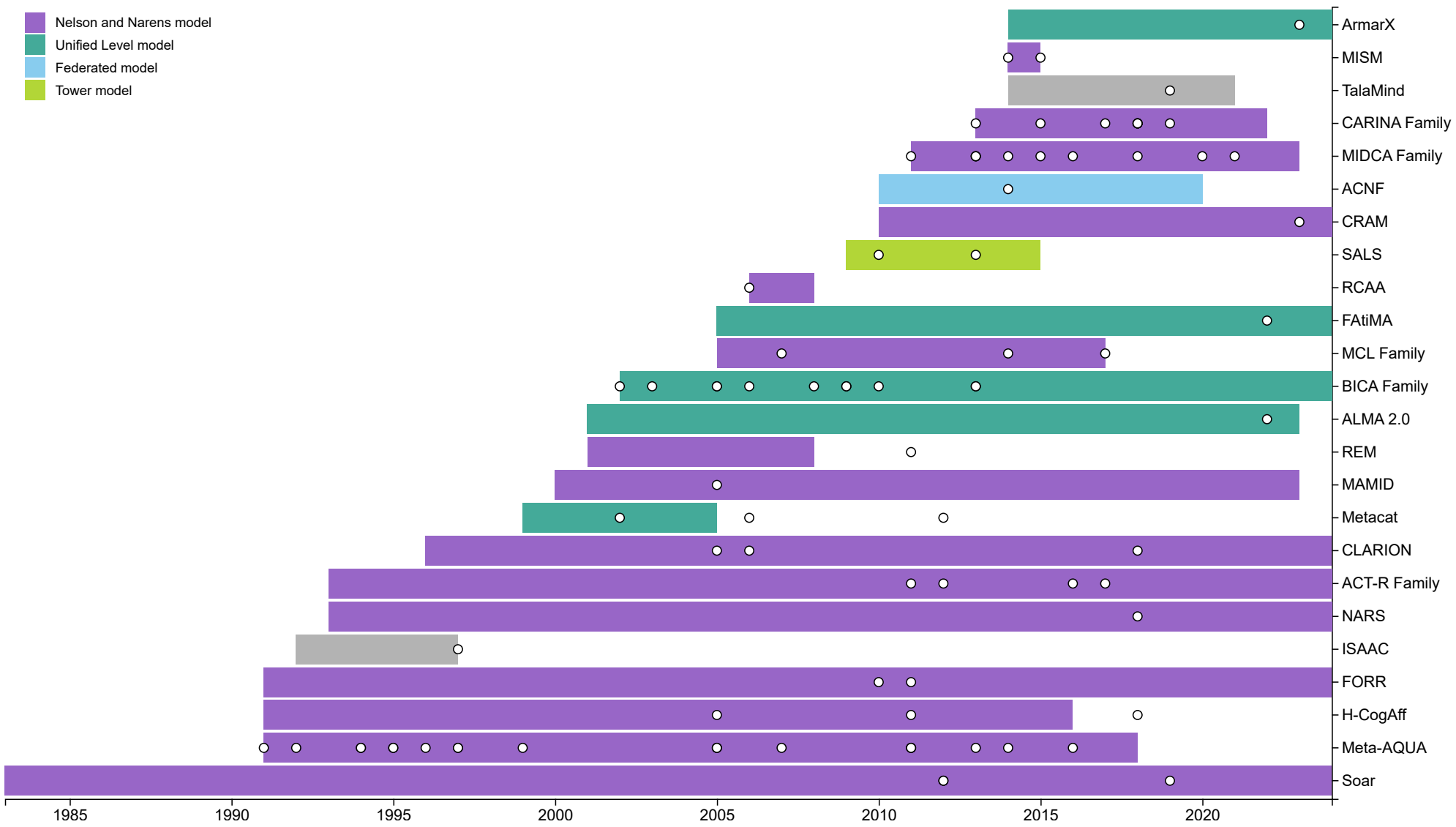

**Fig. 6** Timeline of the development periods for the *named* CMA in this review, color-coded by metacognition model (Figure 10) and inspired by Kotseruba and Tsotsos's (2020, p. 20) Figure 2. Start (end) dates are inferred from the earliest (latest) activity we identified — publications, Git commits, or website updates — because precise dates are not reported. White dots mark publication dates of relevant reports (a single dot may represent several). An interactive version on the companion website lists the sources for every first and last activity

# 4 (A) Model Context

## 4.1 (A1) Used CMAs and (A2) and Publication Dates

We identified 35 CMAs: 26 *named* and 9 *unnamed* (cf. Section 2.6). Figure 5 shows that 11 of the named architectures (42.3%) also appear in the overview by Kotseruba and Tsotsos (2020), along with three further architectures that they mention but explicitly do not analyse. Of those 14, they associated only four with our technique of interest (cf. Section 3.1).

Figure 6 adapts the timeline from Kotseruba and Tsotsos (2020, p. 20 ) for the named CMA in our corpus, visualizing their development windows. Unnamed architectures are omitted because reliable start or end dates are infeasible to identify without labels. The timeline spans more than 30 years (1991–2023). Some architectures — Meta-AQUA, for example — address our topic throughout their entire lifespan. Others, such as FORR and MAMID, feature only isolated relevant reports, suggesting varying importance of the topic.

Figure 7 plots the number of relevant reports per year; activity peaks in the 2010s, yet annual counts remain modest. Together with the mediocre data availability noted in Section 3.2, this reinforces the sense that memories of metacognitive experiences constitute a sideline niche research theme overall, except for some dedicated CMAs. However, because our search query does not cover all terms identified later (Section 4.2), and because our eligibility criteria — notably the focus on explicit cognitive architectures — are rather strict,

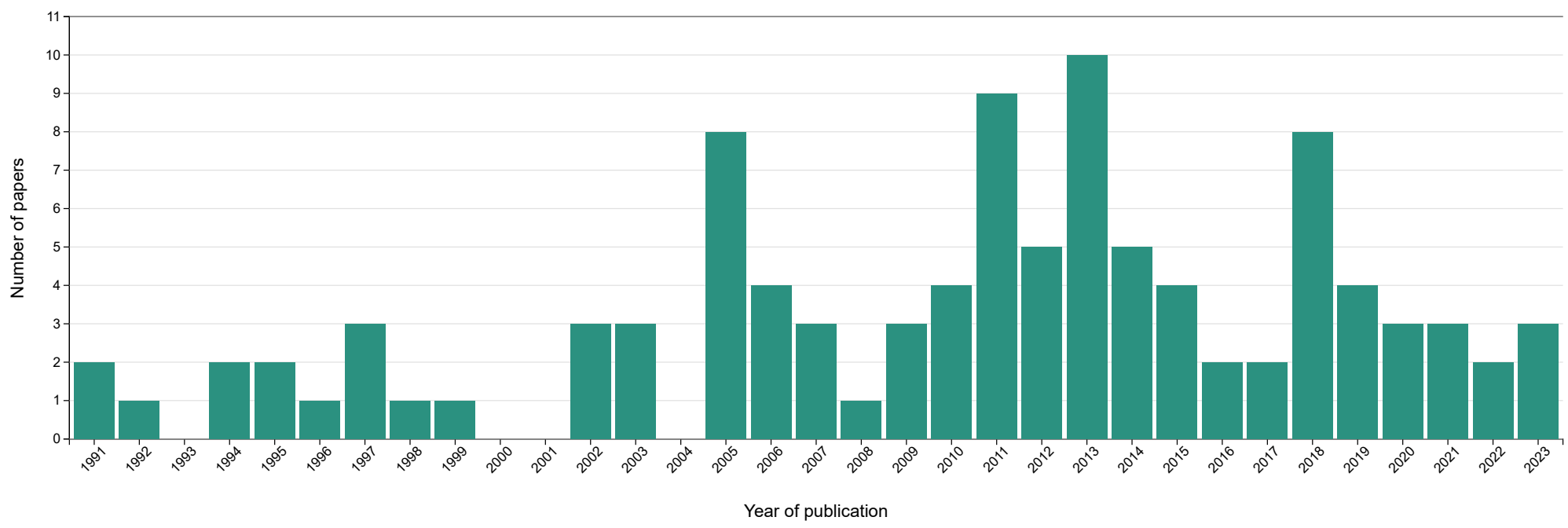

**Fig. 7** Time series plot of the number of identified relevant reports by year of publication. An interactive version is available at the companion website

we are unsure to which extend this impression reflects reality. Related domains outlined in Section 2.1.3 show that similar techniques are common in other contexts as well.

## 4.2 (A3) Terminology

One challenge identified in our initial explorative literature search (Section 2.3) is the inconsistent terminology regarding traces of metacognitive experiences, which complicates direct comparisons across architectures. To impose structure, we summarised — in Table 2 — the most specific terminology each CMA uses for their episodic metacognitive experiences. Some CMAs, such as Soar (Laird 2019), apply no explicit label and instead treat the mechanisms as a natural function of episodic memory; we recorded the generic label nevertheless.

The vocabulary is highly fragmented: Although a few architectures reuse identical phrases — CARINA, for instance, inherits "ReasoningTrace" from its predecessor (Section 4.5) MISM — most terms differ at least slightly, even when generic words like "trace" or "episodic memory" recur. To identify latent themes we semi-automatically grouped the terms by semantic similarity. We start by calculating vector embeddings using *SentenceTransformer* (Reimers and Gurevych 2019) and the model *multi-qa-mpnet-base-cos-v1* (Hugging Face 2021), which is tailored to semantic similarity. We then cluster the vectors with *affinity propagation* (Frey and Dueck 2007) via *scikit-learn* (Pedregosa et al. 2011)[17]. The algorithm yielded 13 clusters (see the raw-data repository), which we manually label by overarching topic.

Our review revealed overlaps and a few misassignments. For example, "traces of the system's inference activity", which the algorithm sorted into the cluster "Processing Trace", but also fits the cluster "Reasoning Trace" due to its emphasis on inferences. Furthermore, BICA's "I-Now/I-Past/I-Meta" denote specific mental states (Samsonovich et al. 2009) and are grouped in the same cluster — but "mental states" is not. We adjusted the clusters accordingly; all changes are documented in the raw data repository. The final set appears in Table 3.

Three clusters revolve around the word "trace", each emphasising a different aspect (problem-solving, processing, reasoning). The clustering algorithm likely split the these because "trace" appears so frequently — an artefact of its explicit inclusion in search query $q$. Oher clusters span technical notions ("Performance Profile", "Processing Trace") through psychological concepts ("Episodic memory", "Experience"). To inform future studies, it is of interest to determine which topics are most prevalent in the relevant literature. We counted the number of distinct architectures — not term occurrences — that use at least one label from each topic. This avoids the pitfalls of automated frequency counts (stemming, part-of-speech ambiguity, bulk mentions in a single architecture, etc.). Figure 8 charts these counts and offers a clear view the terminological fragmentation.

[17]In accordance with the requirements of the chosen model, we use dot-product as the similarity metric.

| CMA | Terminology | Sources |
|---|---|---|
| ACNF | trace of the reasoning | (Crowder et al. 2014, p. 120) |
| ACT-R | representations of mental states, episodic information | (Birlo and Tapus, 2011, p. 116; Reitter and Lebiere, 2012, p. 246) |
| ALMA 2.0 | self-knowledge | (Goldberg 2022, p. 31) |
| AML11 | inference trace | (Arcos et al. 2011, p. 170) |
| ArmarX | meta-information, traces of data through the processing pipeline | (Peller-Konrad et al. 2023, p. 10) |
| BICA | episodic-memory, mental state, I-Now, I-Past, I-Meta, imaginary self | (Samsonovich 2013a, p. 111ff.; 2013b, p. 111ff.; Samsonovich and De Jong 2005, p. 2) |
| BCP13 | immediate processual self, autobiographical data, autobiographical self | (Brody et al. 2013, p. 3) |
| CARINA | self-model, performance profile, internal representation of the reasoning processes, ReasoningTrace | (Caro Piñeres et al. 2019, p. 74f.) |
| CLARION | monitoring buffer | (Sun 2018, p. 29) |
| CRAM | metacognitive experiences, traces, telemetry | (Nolte et al. 2023, p. 319) |
| D20 | replay memory, experience memory replay | (Daglarli 2020, p. 98497) |
| FAtiMA | autobiographical memory | (Mascarenhas et al. 2022, p. 8) |
| FORR | history, trace of each solved problem | (Epstein and Petrovic 2011, p. 44, 51) |
| H13 | operational history, contextualized process performance history | (Helgason 2013, p. 89, 116) |
| H-CogAff | traces, representation | (Kennedy, 2011, p. 235; Sloman and Chrisley, 2005, p. 146) |
| ISAAC | metareasoning representation | (Moorman 1997, p. 40) |
| K10 | reasoning traces | (Kennedy 2010, p. 1) |
| K12 | reasoning trace, episodic memory of mental events, audit trail | (Kennedy 2012, p. 2) |
| K21 | trace | (Kennedy 2021, p. 4) |
| KS02 | trace of execution events, event records | (Kennedy and Sloman 2002, p. 11) |
| MAMID | affective state, feeling of confidence | (Hudlicka 2005, p. 2f.) |
| MCL | episodic memory | (M'Balé and Josyula 2014, p. 206) |
| Meta-AQUA | trace, representational model of [. . . ] reasoning processes | (Cox and Raja, 2011a, p. 7; Gordon et al., 2011, p. 301) |
| Metacat | temporal trace, temporal record, episodic memory | (Marshall 2006, p. 281f.) |
| MIDCA | reasoning traces, episodic memory, representation, self-model, cognitive trace | (Cox et al., 2011, p. 80; Mohammad et al., 2020, p. 4) |
| MISM | ReasoningTrace | (Caro Piñeres et al. 2014, p. 94) |
| MT21 | mental models | (van Ments and Treur 2021, p. 7) |
| NARS | mental sensation, traces of the system's inference activity, internal experience | (Wang et al. 2018, p. 7) |
| PRODIGY | problem-solving trace | (Carbonell et al. 1991, p. 52) |
| RCAA | historical experience, history | (Foltyn et al. 2006, p. 4) |
| REM | trace of processing | (Goel and Jones 2011, p. 154) |
| SALS | reflective traces, representations, experiential event streams | (Morgan 2013, p. 80, 102) |
| SCADS | traces | (Shrager and Siegler 1998, p. 408) |
| Soar | episodic memory | (Laird 2019, p. 225) |
| TalaMind | traces of [. . . ] execution | (Jackson 2019, p. 533) |

**Table 2** Most specific terms (with sources) identified by the screeners for each CMA to denote memories of metacognitive experiences (Definition 3)

| Cluster Topic | Terms |
|---|---|
| **Autobiographical Self** | autobiographical data, autobiographical memory, autobiographical self, imaginary self, immediate processual self, self-knowledge, self-model |
| **Data** | autobiographical data, meta-information, telemetry, traces of data through the processing pipeline |
| **Episodic Memory** | autobiographical memory, episodic information, episodic memory, episodic memory of mental events, experience memory replay, replay memory |
| **Event Records** | event records, experiential event streams, temporal record |
| **Experience** | affective state, experience memory replay, experiential event streams, feeling of confidence, historical experience, internal experience, mental sensation, metacognitive experiences |
| **History** | contextualized process performance history, historical experience, history, operational history |
| **Mental State** | I-Meta, I-Now, I-Past, mental state, representations of mental states |
| **Performance Profile** | contextualized process performance history, monitoring buffer, performance profile |
| **Reasoning Trace** | cognitive trace, inference trace, reasoning trace, trace of the reasoning, traces of the system's inference activity |
| **Representation** | internal representation of the reasoning processes, metareasoning representation, representation, mental models, representational model of [. . . ] reasoning processes, representations of mental states |
| **Problem-solving Trace** | problem-solving trace, reflective traces, temporal trace, trace, trace of each solved problem |
| **Processing Trace** | audit trail, trace of execution events, trace of processing, traces of data through the processing pipeline, traces of [. . . ] execution, traces of the system's inference activity, telemetry |

**Table 3** Semi-automatically generated clusters of the terms from Table 2 based on semantic similarity. Word stems (e.g., trace/traces) are merged for clarity

On average, each term is used by $\approx 5.17$ CMAs (SD $\approx 2.54$). Only the topics "Episodic Memory", "Reasoning Trace", and "Problem-solving Trace" exceed the rule-of-thumb threshold of $mean + 1SD$, so we treat them as comparatively frequent. Yet, even here consensus is weak: merely one-quarter of the 35 CMAs employ a term from the most prevalent category, "Episodic memory". To avoid masking distinct emphases, we initially kept the three automatically generated "trace" clusters separate, although merging them would not be unreasonable. Combined, they would form the dominant category by a wide margin, appearing in $60\%$ $(21/35)$ of the CMAs. At the other end of the spectrum, the topics "Data", "Event Records" and "Mental State" fall below the $mean - 1SD$ and can be considered rarely used.

## 4.3 (A4) Implementation

Over $80\%$ of the named CMAs — but only $30\%$ of the unnamed ones — include a code implementation of the technique, as documented in associated reports; see Figure 9. This difference is statistically significant (Pearson $\chi^2$, $p < 0.001$). A plausible explanation is

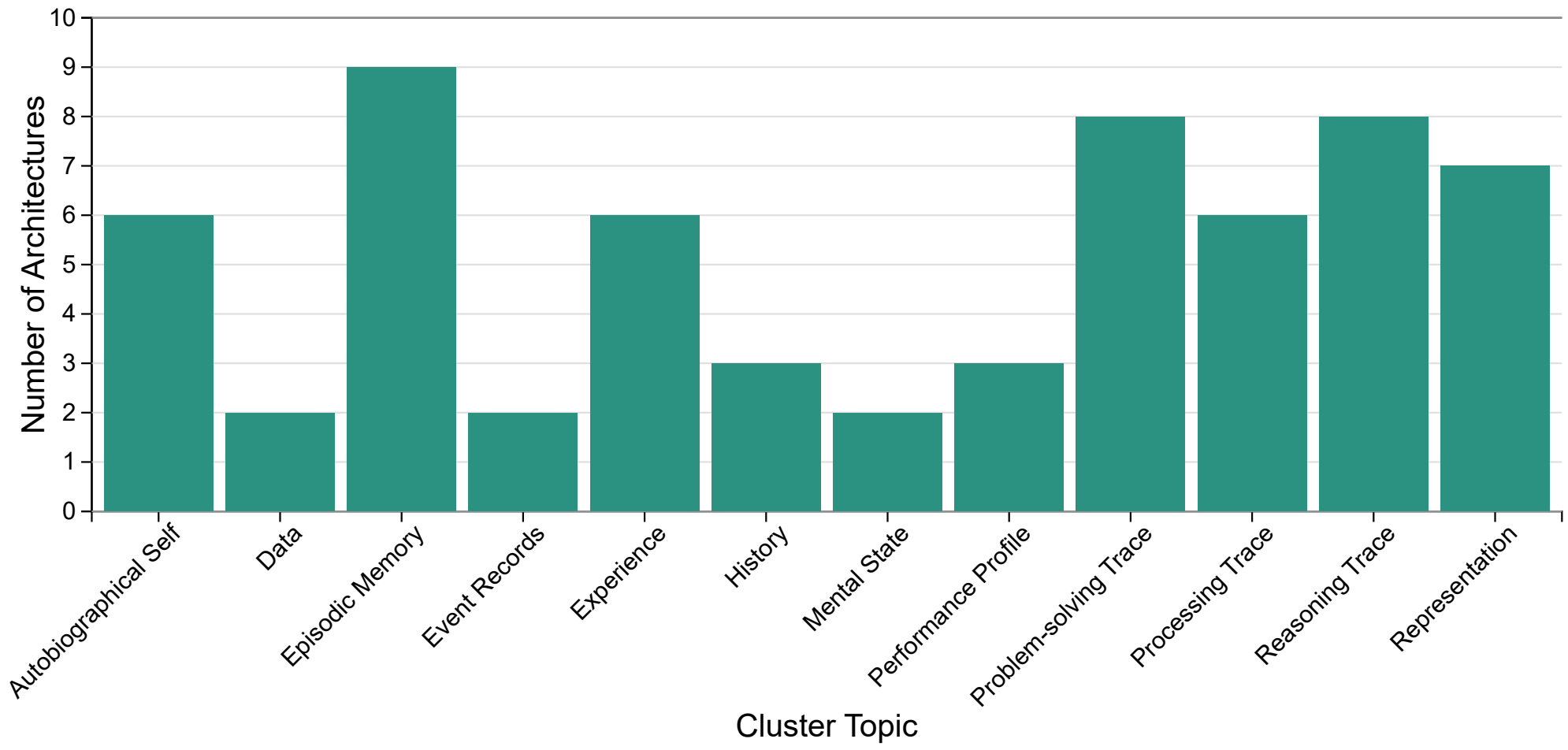

**Fig. 8** Bar chart of the CMA count that use at least one term from each topic (term-topic assignments are from Table 3). An interactive version is available on the companion website

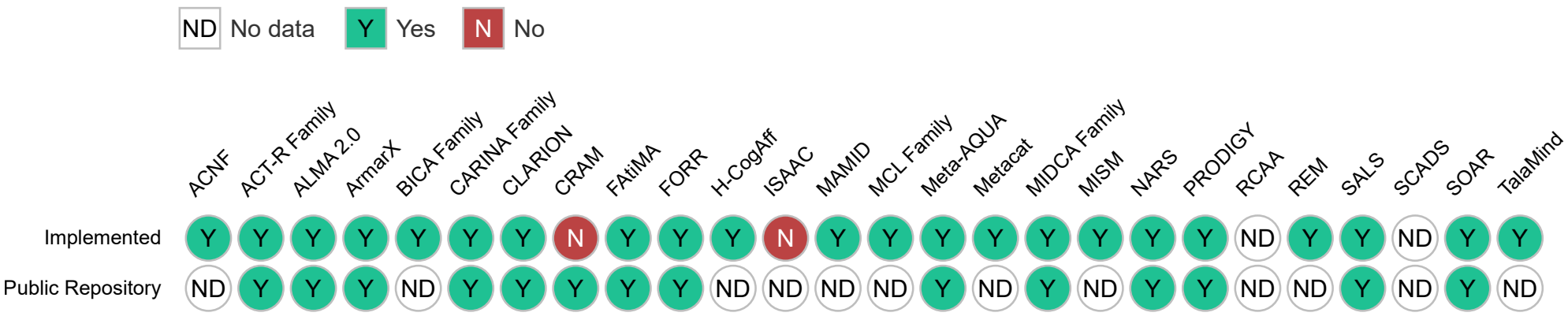

**(a)** *Named* CMAs: top row shows reported implementation; bottom row adds code-repository availability (Git/SVN/. . . ). The companion website links to each repository.

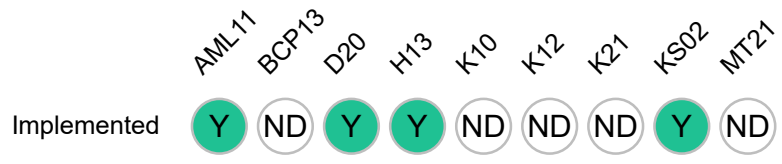

**(b)** *Unnamed* CMAs: implementation status only; no repositories located.

**Fig. 9** Implementation status (A4) for named and unnamed CMAs. Y = yes, N = no, ND = no data

that larger projects, able to support substantial implementation work, tend to adopt explicit names for group identity. Note that a CMA is counted as implemented when at least one paper reports running code; individual ideas within that architecture may still be unrealised.

Because many reports omit details (Section 3.2) open-sourcing the code, e.g., on Git or SVN, greatly enhances reproducibility. We therefore searched for public repositories: web searches for named CMAs and in-text links for unnamed ones. The results appear in Figure 9 and the interactive version links directly to each repository. Repository presence also correlates with longevity and can therefore serve as an indicator of a CMA's maturity: named CMAs with public code remain active for an average of 20.3 years, whereas those without repositories average only $\approx 11.2$ years; no repositories were found for unnamed CMAs, so they are excluded from this calculation.

## 4.4 (A5) Inspiring Non-Computational Models of Cognition

#### *Inspiring Models of Cognition*

Especially for older CMAs such as ACT-R (Anderson 2007), CLARION (Sun 2016), and Soar (Laird 2012), the volume and variety of cited research make it impossible to give a complete account of every inspirations. Still, psychological plausibility emerges as many CMAs, including, e.g., CLARION (Sun et al. 2005). Many CMAs borrow directly from human cognitive functions such as memory and executive control, and some map components onto neural structures: ACNF (Crowder et al. 2014), D20 (Daglarli 2020), and the BICA family (Samsonovich 2013a) all align modules with specific regions of the human brain. BICA, for example, locates metacognitive experiences with episodic memory in the *hippocampus and extrastriate neocortex* (Samsonovich 2013a, p. 111), whereas ACNF and D20 place them in the prefrontal cortex. BICA even leverages its mental state model to explain phenomena such as PTSD and brain lesions (Samsonovich et al. 2009). As notable exception is PRODIGY, which explicitly dismisses any resemblance to human cognition as incidental; instead, its "goal is at reengineering cognition the way it ought to be" (Carbonell et al. 1991, p. 54).

Some works focus on specific aspects of cognition. BCP13 (Brody et al. 2013) uses *active logic* (Anderson et al. 2008) to model the phenomenon of *thick time*, initially conceptualized by Newton (2001) and Humphrey (2006) as the time frame humans perceive as the conscious present. Brody et al. boil this down to the technical implementation of monitoring mental processes during runtime rather than inspecting them retrospectively after completion. Other specialised efforts include K21 [behaviour-change theory (Kennedy 2021)], SCADS [children's strategy discovery (Shrager and Siegler 1998)], FAtiMA [OCC model of emotions (Ortony et al. 1988; Mascarenhas et al. 2022)], and ArmarX [*Multi-Store Model* (Atkinson and Shiffrin 1968; Peller-Konrad et al. 2023)].

#### *Inspiring Models of Meta-Cognition*

The CMAs disagree on whether storing, retrieving, and processing of episodic metacognitive experiences is an automatic, unconscious procedure or deliberate. Soar, for instance, stores episodes automatically (Laird 2012, p. 229), whereas Metacat decides which events merit storage (see Section 5.3). Conversely, Soar determines when and which episodes should be retrieved via its typical operator-selection mechanism (Laird 2012, p. 230), while Metacat's pattern clamping appears to occur automatically, as do Meta-AQUA's *Meta-Explanation Pattern* (Meta XP) matching and MIDCA's metacognitive cycle (see Section 6.3).

It is clearer how CMAs architecturally structure the interaction between cognition and metacognition. Different psychological models (Figure 10) disagree as to whether there is only one meta-level, several — and how they interact —, or just a singular unified system (Zheng et al. 2023). We classify the CMAs by the employed model as shown in Figure 11. A majority ($\approx 57.1\%$) of CMAs, including sophisticated frameworks such as the MCL Family, MIDCA and MISM, adopt the the early Nelson and Narens (1990) model (Figure 10a), separating object-level and meta-level. The MCL Family's younger members, e.g., GPME, are designed as external add-on meta-levels to existing CCAs that themselves only feature object-level cognition (Schmill et al. 2007; M'Balé 2017). MIDCA, being highly inspired by the MCL Family

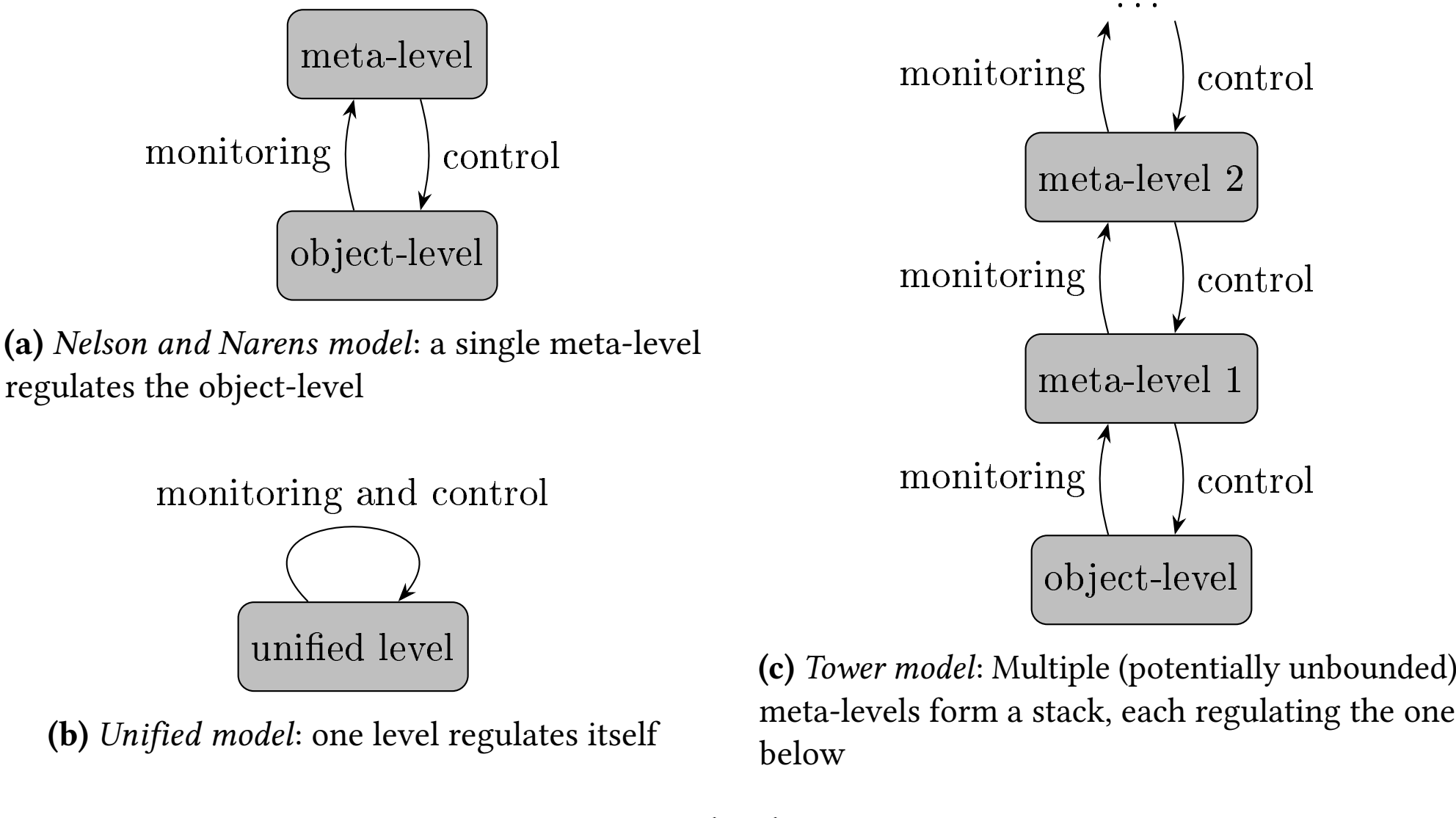

**(a)** *Nelson and Narens model*: a single meta-level regulates the object-level

**(b)** *Unified model*: one level regulates itself

**(c)** *Tower model*: Multiple (potentially unbounded) meta-levels form a stack, each regulating the one below

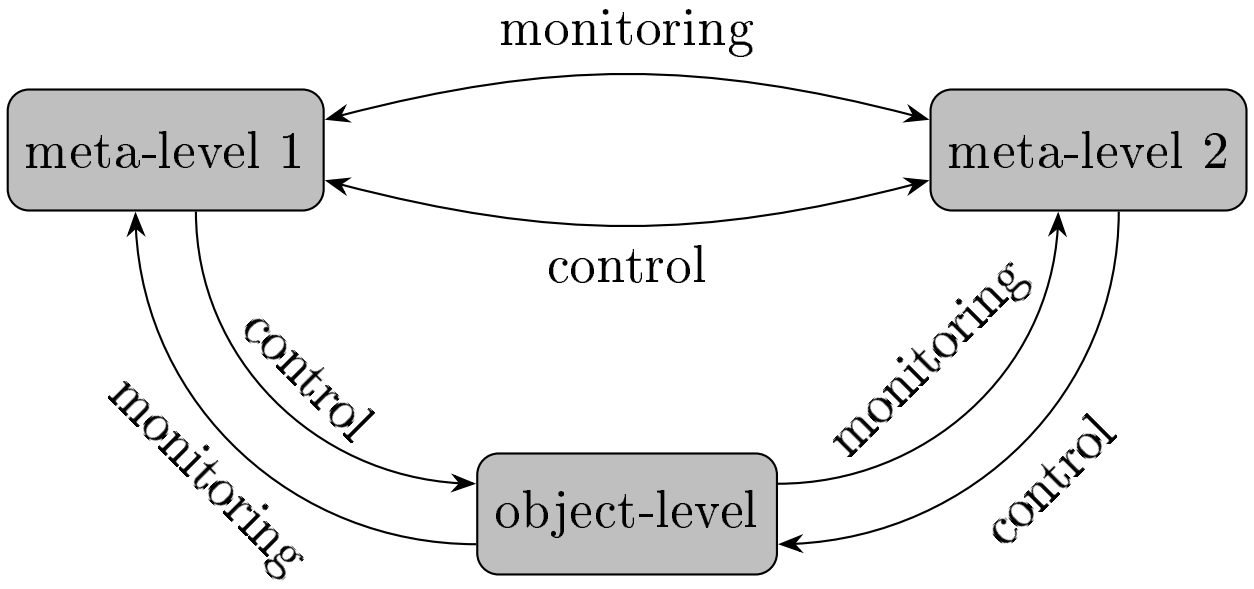

**(d)** *Federated model*: Multiple peer meta-levels run in parallel, each regulating the object-level(s), and one another; only two meta-levels are shown for clarity, but more are possible

**Fig. 10** Schematics of the four principle meta-level models in the surveyed CMAs

[Cox et al. (2011); see Section 4.5], also emphasizes this distinction by identifying itself as a "dual-cycle architecture", with one cycle dedicated to metacognition (Cox et al. 2016). Note that these systems often mirror object-level patterns at the meta-level, For example, MIDCA's meta-level cycle repeats the same phases as its object-level cycle (Cox et al. 2016, pp. 3713), and Meta-AQUA reuses its story understanding algorithm to explain traces of its metacognitive experiences (Ram and Cox 1995, p. 221); more on this later in Section 6.3. Besides the apparent advantage of simplifying the development and maintenance of CMA software, this suggests that many researchers believe metacognition and cognition to be closely related.

Nearly a fifth of CMAs reject a distinct meta-level and process meta-representations via the very mechanisms that handle object-level cognition; Figure 10b shows such a *unified level*. The authors of BICA even posit that if different modules are responsible for cognition and metacognition, one may not speak of true metacognition because cognition would not be turned on itself (Samsonovich 2009). Nevertheless, the CMAs in the BICA family still

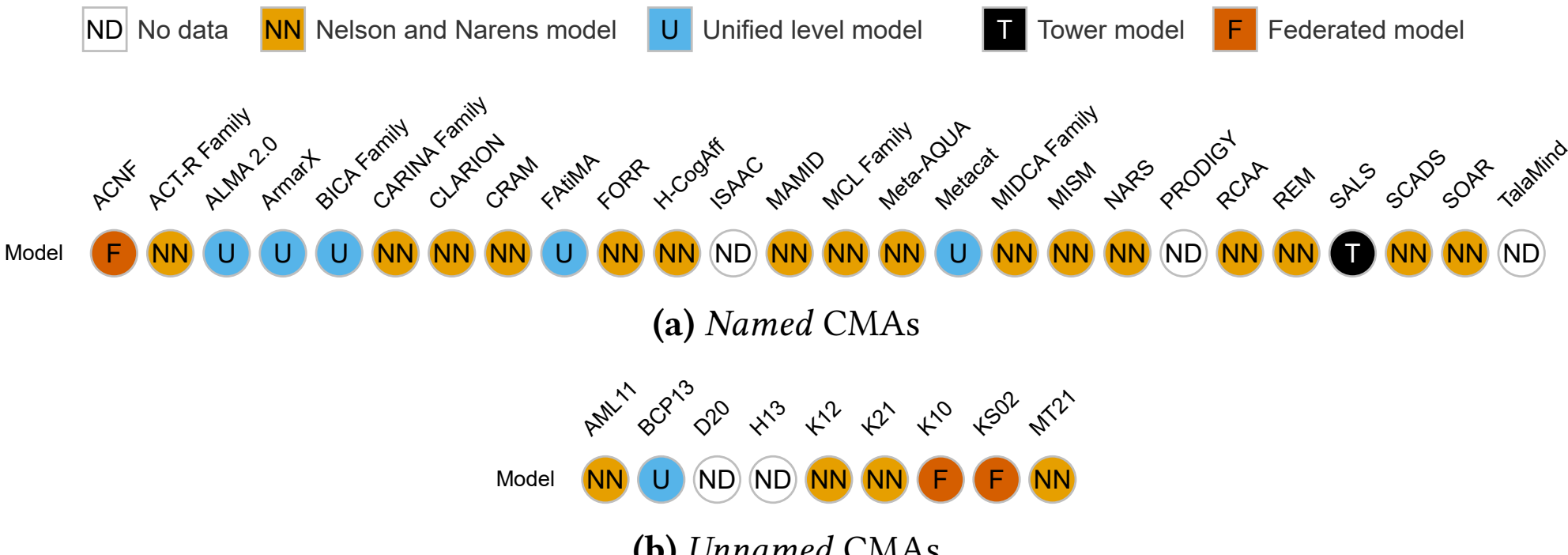

**(a)** *Named* CMAs

**(b)** *Unnamed* CMAs

**Fig. 11** Meta-level models adopted by each CMA. NN = Nelson and Narens model (Figure 10a), U = Unified model (Figure 10b), T = Tower model (Figure 10c), F = Federated model (Figure 10d), ND = no data. Few architectures state their model explicitly; Screener A inferred classifications, so these data were not considered in Figure 3

keep a mental state "I-Meta" separate from "I-Now" until metacognition occurs, for which "I-Meta" briefly assumes the role of "I-Now". Although a single component processes both, BICA neatly separates the respective data. Advocates for a unified level often cite a familiar *Quis custodiet ipsos custodes?* argument: if metacognition requires a module separate from cognition, yet another module is required for meta-metacognition, and so forth. For instance, a meta-meta-level could balance metacognition's computational overhead with any achieved performance gain. Recursively applying this principle yields a theoretically infinite *tower* of stacked meta-levels (Kennedy and Sloman 2002), as depicted in Figure 10c. Since human meta-metacognition is empirically supported (Recht et al. 2022), such rationales are not restricted to thought experiments. One reviewed CMA, SALS, indeed features five stacked layers, each monitoring and controlling the lower one, starting with the lowest *Built-in Reactive Layer*, through a *Super-Reflective Layer* (Morgan 2013). CMAs following Nelson and Narens' (1990) simple model deliberately ignore meta-metacognition.

A fourth strategy — the *federated model* (Figure 10d) — appears in KS02 (Kennedy and Sloman 2002, 2003; Kennedy 2003), and, in spirit, ACNF (Crowder et al. 2014) and K10 (Kennedy 2010). Multiple peer meta-levels run in parallel, each monitoring the object-level (possibly also multiple) and its fellow meta-levels. If one of these meta-levels fails or is compromised, the others can diagnose and repair it.

Other philosophical accounts, such as Fletcher and Carruthers's (2012) theory that metacognition emerges from interactions among reasoning systems rather than from a dedicated layer, are not implemented in any CMAs surveyed[18].

## 4.5 (A6) Related CCAs and (A7) Main Innovations to those

We distinguish four types of architectural lineages:

[18]The only potential exception is TalaMind, which adopts a *Society of Minds* approach: it lacks a dedicated meta-level yet can still reflect on past thoughts. The sole relevant record (Jackson 2019) provides no further details; we instead refer to Jackson (2014).

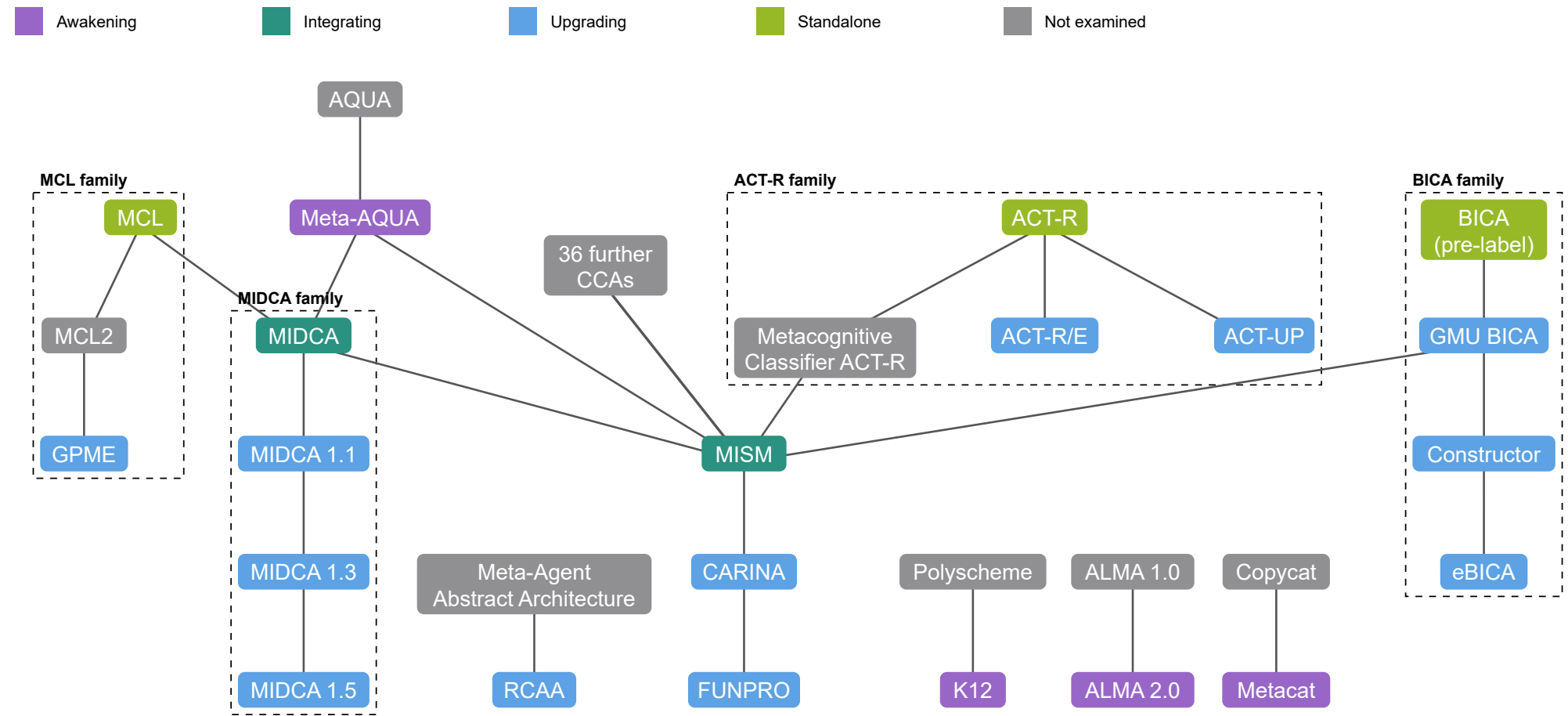

**Fig. 12** Lineages of CCAs that have identifiable predecessors or successors. Ancestors appear above their descendants; dashed boxes enclose summarized *families*. To conserve space, 36 predecessors of MISM are omitted, (five of them are otherwise relevant but have no additional relatives; see Section 4.5). Nodes are color-coded by category

*Standalone CMAs* lack explicit parents. They borrow only general ideas from earlier CCAs and develop their algorithms and representations independently.

*Awakening CMAs* extend a parent CCAs that possessed little to no metacognitive capability, adding a holistic implementation of metacognition.

*Upgrading CMAs* evolve a single parent CMA by refining its existing mechanisms and introducing new ideas. This review often summarizes closely related upgrading lineages as a *family* (e.g. the MIDCA family includes versions 1.1, 1.3, and 1.5).

*Integrating CMAs* intermix elements from several parent CMAs, each of which model different aspects of metacognition, to create a unified and encompassing framework.

Figure 12 traces the lineages of awakening, upgrading and integrating architectures, and summarizes the families we identify. Although many reports compare newly introduced CMAs with existing ones, only 38.5% of the named architectures actually inherit principles, algorithms, and representations from a predecessor (at least as reported in the relevant papers). When unnamed architectures are included — and removing the four awakening CMAs whose parents lacked metacognition (Metacat, K12, Meta-AQUA, and ALMA 2.0) — the share falls to 20.0%. Note that this number accounts for the architectures that genuinely build upon existing metacognitive principles rather than introducing them outright. For instance, Metacat extends Copycat (Mitchell 1993) with episodic memory and metacognitive abilities, which Copycat both lacked (York and Swan 2012). Likewise, Meta-AQUA reapplies AQUA's (Ram 1989) story explanation system to diagnose its own reasoning and devise compensatory learning strategies (Ram and Cox 1995).

#### *Integrating CMAs*

Only MIDCA and MISM qualify as *integrating* CMAs. MIDCA inherits the object-/meta-level split from the MCL Family (Cox et al. 2011) and embeds Meta-AQUA's self-explanation representations and algorithms into its meta-level (Cox et al. 2016). How the latter is technically achieved is unclear: Meta-AQUA's *Meta XP matching* — used to detect reasoning issues — resembles matching graph queries against traces of metacognitive experiences (Ram and Cox 1995); see Section 6.3. MIDCA, however, represents traces as linked lists (Cox et al. 2016) or ordered dictionaries (Mohammad et al. 2020) that record the inputs and outputs phases such as planning. The relevant reports do not document how Meta XP matching translates to these simpler, less expressive structures.

MISM occupies a different niche: it is a metamodel distilled from 40 metacognitive models and architectures (Caro Piñeres et al. 2014). Key concepts and relations were extracted from a random half of those architectures, merged into a draft, and validated against the remaining half. Thus MISM incorporates elements from many CMAs reviewed here — Meta-AQUA, CLARION, MIDCA, Soar, MCL, MAMID, BICA, REM, H-CogAff — and from an ACT-R variant (Vinokurov et al. 2011) outside this review's scope. By design it is abstract: While MISM covers concepts such as "ReasoningTrace", "Memory", and "MonitoringTask" (Caro Piñeres et al. 2014, p. 94), it omits algorithmic details, offering instead UML templates that have guided implementations such as CARINA (Caro Piñeres et al. 2018) and FUNPRO (Caro Piñeres 2015).

#### *Upgrading CMAs*

We classify ALMA 2.0, CARINA, FUNPRO, K12, RCAA, and the descendants of the ACT-R, BICA, MCL and MIDCA families as *upgrading* CMAs. MIDCA itself has progressed through versions 1.1 (Cox 2013), 1.3 (Cox et al. 2016), and 1.5 (Cox et al. 2021), each implementing concepts absent from its initial design. Other families branch more clearly: For instance, the descendants of the ACT-R and BICA families focus on different aspects, e.g., eBICA adds a model of emotions (Samsonovich 2013a), and ACT-R/E extends ACT-R towards embodiment in robots (Birlo and Tapus 2011). CARINA is a direct instantiation of MISM, and FUNPRO adapts CARINA for intelligent tutoring (Caro Piñeres et al. 2019). The relevant reports do not clarify how RCAA improved upon its predecessor.

# 5 (B) Data Content and Structure

## 5.1 (B1) Data Type

In their overview, Kotseruba and Tsotsos (2020, pp. 14–28) classify CCAs according to the kind of data they process — *symbolic*, *sub-symbolic*, or *hybrid*. Because few architectures provide self-assigned labels, Kotseruba and Tsotsos apply explicit definitions: symbols are atomic building blocks of syntactically and semantically meaningful expressions for reasoning; sub-symbolic data is associated with the metaphor of machine learning, especially neural networks; hybrid data combines both. The relevant reports are similarly sparse, so we reuse these definitions to address question (B1). Note that (B1) focuses only on the data type of metacognitive experiences. Moreover, more than half of the CMAs we analyse are absent from Kotseruba and Tsotsos' survey (Section 4.1), so our study is not a mere replication.

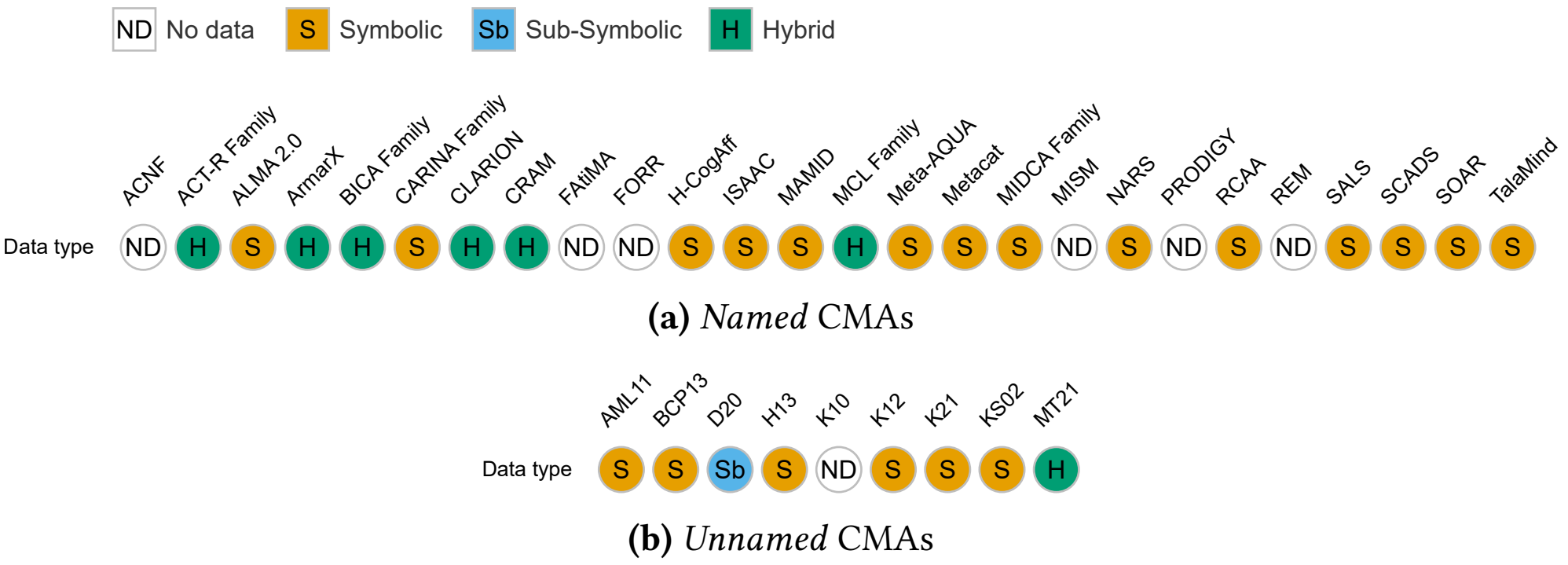

**(a)** *Named* CMAs

**(b)** *Unnamed* CMAs

**Fig. 13** Data types used by the CMA. S = symbolic; Sb = sub-symbolic; H = hybrid; ND = no data. Few architectures explicitly specify the data type they employ; the rest are classified according to the Kotseruba and Tsotsos's (2020) definition (cf. Section 5.1)

As detailed in Figure 13, the majority ($\approx 56\%$) of the CMA represent metacognitive experiences symbolically, about $22\%$ employ a hybrid approach, and only one architecture ($\approx 3\%$) relies exclusively on sub-symbolic data. Section 5.2 details the symbolic cases; here we highlight how hybrid and sub-symbolic solutions differ from them. Hybrid architectures simply attach sub-symbolic metadata to symbolic traces — utility values [ACT-R (Stevens et al. 2016)], indexing metrics [BICA (Samsonovich 2006), GPME (M'Balé and Josyula 2014), MT21 (Canbaloğlu and Treur 2023)], and confidence scores for candidate solutions [CLARION (Sun 2018)]. ArmarX (Peller-Konrad et al. 2023) and, in principle, CRAM (Nolte et al. 2023) log metacognitive experiences as discrete facts, e.g., about module outputs, and link them to sensor streams and higher-level reasoning events, thereby explaining how each symbolic event arises from sub-symbolic data. The only purely sub-symbolic CMA is D20 (Daglarli 2020), which stores raw states, state transitions, and reinforcement rewards.

## 5.2 (B2) Content

Because the data collected for Item (B2) are highly heterogeneous, we group the content of the different CMA's metacognitive experiences by theme. The following architectures — K10, MISM, PRODIGY, REM, and TalaMind — provide too little detail to include.

#### *Events, States and their Types*

As we focus on metacognitive experiences with a temporal character, it is not surprising that most data are associated with the notion of a *mental event* (some authors might use other terms), that reifies not only a time segment during which the CMAs process information but also a distinct cognitive activity, especially in systems featuring parallel processing. GPME, for example, merges everything its host CCA (i.e., its object-level) exposes at one "moment" into a single frame (M'Balé and Josyula 2014; M'Balé 2017). Though we go into more detail in section Section 5.5, nearly all CMAs label such events with a type — domain-specific, implementation-specific, or lent from psychology — e.g., "Change letter-category of rightmost letter to successor" (Marshall 2006, p. 284), "memorizing" (Nolte et al. 2023, p. 326), or "copySMUtoBCPUinput" (Flórez et al. 2018, p. 624). Of course, the types vary heavily between different CMAs. GPME is the lone exception; it tags events only with sub-symbolic fingerprints plus metadata (M'Balé and Josyula 2014; M'Balé 2017).

Events that signal "failure" (CRAM, Meta-AQUA, RCAA), as "snags" (Metacat), "anomalies" (CARINA, the MCL Family), and so forth — might seem as prime candidates for triggering error handling (see Section 6.1). Yet almost no CMA represent them in episodic memory — in fact, Metacat (Marshall 2002) is the only one we know. More often, failures are detected *via* the representations in episodic memory to initiate countermeasures (cf. Section 6.3).

Even the purely sub-symbolic D20 (Daglarli 2020) encodes its metacognitive experiences as events that correspond to the steps of an adapted *Q-learning* loop [see, e.g., (Jang et al. 2019)]. Tuple $(s, a, r, s')$ are stored in an *experience memory replay* and capture successive states $s$ and $s'$, the object-level's intervening action $a$, and the reinforcement reward $r$. Unlike standard Q-learning, however, $r$ is not imposed from the outside after each action, but is determined by D20's meta-level, which is in line with our definition of metacognitive experiences. Various factors, such as past actions and stimuli, detected errors, and "dopaminergic gain", are taken into account for computing $r$ (Daglarli 2020, p. 98499).

Two architectures — MT21 and the BICA family — record metacognitive experiences as *mental states* rather than events. Samsonovich and De Jong (2005, p. 2) conceptualizes these as "discrete instances" of BICA's self that embed schematic event descriptions such as "Verify hypothesis: Worried about wife; Answer: No" (Samsonovich 2009, p. 128). Unlike other CMAs who tacitly attribute their metacognitive experiences to their retrospective, BICA models different perspectives through the mental states. Each state bears a perspective prefix, (*I*, *She*, etc.) and an attitude suffix (*Now*, *Meta*, *Imagined*, etc.), producing labels such as *He-Now-She-Now* — BICA's simulation of "his" current belief about "her" current state. BICA also considers egocentric states, e.g., its own conscious self *I-Now*, *I-Imagined* of imaginary experiences such as dreams (Samsonovich and Nadel 2005), and *I-Meta*. Independent from BICA, Castelfranchi and Falcone (2019) argue that modelling the partner's views of an agent is vital for successful interaction: if the human misinterprets a statement or action, the CMA can only identify this misunderstanding by comparing its own metacognitively monitored state with how the human models the CMA.

Still, the BICA family reifies transitions between states as events, although inexplicitly: In an example from (Samsonovich et al. 2009, p. 118), *I-Now* holds the intention to drive, and *I-Next* encodes the expected post-drive state[19]. During execution, *I-Now* is continuously updated (e.g., current distance to the goal) and archived in episodic memory for later replay (Samsonovich et al. 2009, p. 119).

The developers of CRAM's metacognitive experiences partly share BICA's stance and argue that "any characterization of an objective occurrence unexceptionally depends on the observers' subjective narrative" (Nolte et al. 2023, p. 323). However, instead of representing these in different knowledge graphs (such as BICA's mental states), CRAM reifies them according to its ontological commitments (see Section 5.4). For that, CRAM associates *Event Type*s to *Event*s defined by an observer's *Narrative*; e.g., CRAM itself might interpret some *Mental Action* as *Decision Making*. CRAM likewise represents the roles of event participant depending on its own perspective.

[19] Although driving itself is a ground-level activity, processing the meta-representation of the intention to drive is treated as metacognition.

Originally, BICA's mental states were purely representational and required external "drivers" to be processed and modified (Samsonovich and De Jong 2003, p. 1031). Later versions grant some states — chiefly *I-Meta* — the agency to modify others directly (Samsonovich et al. 2009, p. 117). States may also change attitude; for example, *I-Now* becomes *I-Past* once a significant event occurred or sufficient time has passed (Samsonovich et al. 2009, p. 119). In summary, BICA's metacognitive experiences are distributed across numerous mental states and embedded in a more encompassing simulationist *Theory of Mind* (ToM, cf. Apperly 2011) approach (Samsonovich et al. 2009)[20].

RCAA models both events, e.g., to later analyze their frequency, but also the state of the working memory as a "very coarse digest of data from the blackboard" (Foltyn et al. 2006, p. 4).

### *Input and Output*

Logged events are typically associated with rich metadata, especially the input they consume and output they produce. ACT-R (Prezenski et al. 2017), ArmarX (Peller-Konrad et al. 2023), Meta-AQUA (Ram and Cox 1995, p. 221), Metacat (Marshall 2006), SALS (Morgan 2013), and GPME (M'Balé and Josyula 2014) all tie logged events to a symbolic representation of their main results. These results vary widely based on the problem domain and the CMA general functioning: GPME, for example, tracks both a raw image as processed by its host and the faces that the host detects in it (M'Balé and Josyula 2014). Metacat associates input problems with identified *themes*, e.g., $abc \Rightarrow abd; xyz \Rightarrow ?$ with $a$ *AlphabeticPosition opposite* $z$ (Marshall 2006, p. 282). Other architectures record even richer context. ACNF links the "initial state of the problem situation (the contents of the focus, relevant coalitions of codelets, and feature values of relevant concepts, relevant registers in working memory, etc.)" with intended goals and action plans (Crowder et al. 2014, p. 119). H13 likewise tracks internal processes together with any achieved goal states (Helgason 2013).

CARINA logs extensive details for every cognitive functions: input and output parameters, pre- and post-conditions, priority, activation status, goals and sub-goals, and the sequence of visited states (Flórez et al. 2018, pp. 622–623). Besides logging the input and output of cognitive phases [(Cox et al. 2016); cf. Section 6.3], MIDCA interleaves *mental actions*, of which there are eight types, with *mental states* such as "Selected Goals, Pending Goals, Current Plan, World State, Discrepancies, Explanations, [and] Actions" (Dannenhauer et al. 2018, p. 6). CRAM represents parameters like *Input*, *Output*, *Premise*, *Conclusion*, *Sender*, *Receiver*, *Performer* as participant roles within reified events (Nolte et al. 2023). SALS links its *change events*, which only concern changes to the knowledge base, to the specific bits of information added or removed (Morgan 2013).

### *Justifications*

Some CMAs also log decision justifications, expressed either as logical arguments or as heuristic-contribution scores [e.g., FORR (Epstein and Petrovic 2011, p. 44)]. Rule-based systems sometimes log all rules fired — or even the entire inference chain — on the way to the result. Examples include KS02 (Kennedy and Sloman 2002), AML11 (Arcos et al. 2011), H-CogAff (Kennedy 2011), K12 (Kennedy 2012), BCP13 (Brody et al. 2013), CARINA (Flórez et al. 2018), K21 (Kennedy 2021), and ALMA 2.0 (Goldberg 2022). Especially logic-focused systems do not explicitly abstract such inference chains to an event level; for them, they directly represent the traced metacognitive experiences, whereas for others, they would fall under the event-associated main results mentioned above. *Operator*-based systems log the operators they apply for the same reason, as in SCADS (Shrager and Siegler 1998) and Soar (Laird 2019). We treat each recorded inference or operator application as an event in the sense defined above — albeit at very fine granularity.

[20]There is ongoing debate about whether (human) metacognition emerges from agents applying ToM to themselves. We refer to Proust (2007) for an account on animal experiments addressing this question.

#### *Arousal, Emotions, and Feelings of Confidence*

In some cases, the associated metadata also describe a CMA's arousal [ACNF (Crowder et al. 2014); GPME (M'Balé and Josyula 2014)], emotions [ACNF (Crowder et al. 2014); FAtiMA (Mascarenhas et al. 2022); GPME (M'Balé and Josyula 2014); NARS (Wang et al. 2018)], and feelings of confidence [CARINA (Flórez et al. 2018, pp. 622–623); CLARION (Sun 2018); K12 (Kennedy 2012); MAMID (Hudlicka 2005)] that accompanied a mental event.

#### *Start and End Times*

Systems represent the temporal dimension in two broad ways. *Discrete-instant models* treat time as an ordered sequence of ticks: BCP13 increments an index for every inference step (Brody et al. 2013), and KS02 tallies its processing cycles (Kennedy and Sloman 2002). *Metric models* record real clock time: MIDCA logs the start of each processes (Dannenhauer et al. 2018); ACT-R (Reitter and Lebiere 2012) and CARINA Flórez et al. (2018, pp. 622–623) capture both start and end times. SALS (Morgan 2013) also timestamps events, while NARS (Wang et al. 2018) stores only a delta of the elapsed interval between successive events.

#### *Sequence Structure*

Without parallelization, chronological traces may be understood as a causal chain — as appears to be the case for Metacat, Soar, KS02, and SCADS. Meta-AQUA makes causality explicit: trace nodes representing states or processes are linked via relations such as *results*, *enables* or *initiates*, each with defined domains and ranges (Ram and Cox 1995, p. 218). For instance, the state "Run out of gas" *enables* the process "Infer", which then *results* in the state "Should have filled up with gas when tank was low" (Cox 2011, p. 144). These links are constrained by temporal order (Cox and Ram 1992, p. 124) and allow Meta-AQUA to encode complex, non-linear narratives in graph form.

The programming language Funk2, underlying SALS, also traces computational events as graphs, treating parent events as causes of their children (Morgan 2010). CRAM follows a similar approach to Meta-AQUA, linking events by temporality and causality, but making the latter explicit via a small taxonomy of causality relations. For instance, a *Memorizing* event can *terminate* a *memory-containment* state in which some information was previously unknown (Nolte et al. 2023, p. 326). Designers of both Funk2 and CRAM argue that such graph structures are necessary to represent parallel information processing in (Morgan 2010; Nolte et al. 2023). One difference is that CRAM links memory states directly to events, whereas SALS manages the two separately at every layer (Morgan 2013, p. 80).

GPME (M'Balé 2017, p. 137–138) weaves temporal and weak causal links between events and adds spatial links (co-location) plus compositional links that form an event hierarchy, though the purpose of this hierarchy is not explained. RCAA likewise "explicitly care[s] about the order of events as they appeared" to infer causality (Foltyn et al. 2006, p. 4), but it is unclear whether its representation is linear or graph-based. Finally, visualizations suggest that H13 records a tree-shaped trace (Helgason 2013, p. 115, 138).

#### *Absent Data*

The absence of events can be informative as well: After learning typical behaviour, KS02's (Kennedy and Sloman 2002) meta-level identifies anomalies when expected events fail to occur (see Section 6.3). Instead of retrospectively searching for missing events, Meta-AQUA explicitly assigns non-monotonic truth values to trace nodes — e.g., the node representing some information that could not be retrieved is tagged "*out of the current set of beliefs*" (Cox and Ram 1991, p. 225).

Some CMAs intentionally omit specific metacognitive experiences from episodic memory. Soar drops mental imaginary, substates that are reconstructable from other included data, and episodic memory retrieval events, which would otherwise "lead to extremely large episodes" (Laird 2012, pp. 228–230). BICA, by contrast, keeps an unlimited episodic memory to remember every lived experience (Samsonovich 2006).

#### *Peculiarities of Specific CMAs*

K12 (Kennedy 2012) is supposed to log four parallel traces, each concerning different metadata: (1) the CMA's knowledge and perception plus confidence in ensuing inferences; (2) objectives and options, together with justifications for accepting or rejecting each option; (3) the focus of attention and ideas generated; and (4) the "history of [...] disruptive events" and "history of changes in subjective difficulty of a task due to other pressures" (Kennedy 2012, p. 3). The rationale for keeping these streams separate is not explained.

## 5.3 (B3) Level of Abstraction

Because most papers say little about the granularity or abstraction of their metacognitive experiences, we could at best infer them from examples, which risks error. We therefore limit this section to the reproduction of unambiguous information.

As with any logging system, trade-offs exists between log resolution (how many steps are recorded) and utility. CCAs could, unlike humans, keep an exact and complete memory, which might avoid problems such as false memories or forgetting. However, very high data resolution inflates storage and analysis cost[21]. Conversely, too little resolution impairs improvements from trace analysis. The literature features two common mitigation strategies:

*Aggregation:* low-level events are summarized into more abstract events.
*Selective logging:* only events deemed relevant are recorded; the rest are discarded.

Resolution and abstraction, as used here, are independent: a highly abstract trace can still log either most events (high resolution) or only a filtered subset (low resolution). Because the studied CMAs differ fundamentally, quantitative comparison is impractical, but many authors discuss the issue qualitatively.

CMAs with the highest-resolution traces log every reasoning step. Soar, for example, can record each processing cycle, or just those leading to external action (Laird 2012, p. 229). In either case, Soar's episodic memory is low in abstraction: like BICA's (Samsonovich 2013a), it simply stores working memory snapshots. Similarly, KS02 logs every rule fired, plus checks of their preconditions, and ALMA 2.0 traces each individual inference step (Goldberg 2022).

[21] Zilberstein (2011) offers a general analysis of the computational overhead introduced by CMAs employing metacognition.

Metacat, by contrast, records only events that exceed an (unspecified) importance threshold (Marshall 2006). The intuition given is that out of hundreds of events, Metacat deems only a few dozen relevant enough to record. Out of hundreds of events it keeps only a few dozen — chiefly the main ideas and "snags" it encounters during analogy detection.

H13 also limits episodic-memory size, deemed otherwise unmanageable (Helgason 2013). Instead of storing only critical events, it ontologically generalizes goals and internal processes: instances map to classes (e.g., 3.0 to *Number*), classes to superclasses (e.g., *Chair* to *Physical_Object*), producing patterns that future events can match (Helgason 2013, pp. 110-119, 135). These abstract patterns help identify processes that proved useful previously.

According to Kennedy (2012, p. 2), K12's traces "represent mental states on a high level, and do not include the computational 'fine structure' "; a mapping can recover details when needed. Cox (1997a) similarly argues that, as in humans, "representation should have a level of detail that reflects the causal structure and content of reasoning failures" (Cox 1997a, pp. 215–216); Meta-AQUA's traces are shaped accordingly. CRAM, inspired by software tracing tools, supports flexible abstraction: complex (mental) events may be composed of simpler ones (Nolte et al. 2023, pp. 320,324–325). For instance, a discourse between executive functions and memory decomposes into queries and responses, and again into comprehending and answering incoming messages, and so forth. Funk2, the language underlying SALS, also allows flexible abstraction, e.g., machine-code tracing can be disabled (Morgan 2010).

## 5.4 (B4) Formal Specifications

Only 13 of the 35 CMAs examined ($\approx 37.14\%$) rely on schemas, formalisms, or precise descriptions according to the relevant reports to represent their metacognitive experiences.

Some authors adopt mathematical notations. MIDCA (Dannenhauer et al. 2018; Cox et al. 2021) and CARINA (Flórez et al. 2018) use notation reminiscent of denotational semantics to formalize trace structure. In CARINA these symbols mainly label and organizing trace components rather than convey semantics. Much stricter, Helgason (2013, p. 13) proposes a compact grammar defining the syntax of valid patterns that H13 abstracts from individual events (see Section 5.3). As mentioned in Section 5.2, Daglarli (2020) models D20's metacognitive experiences as tuples of sucessive states, actions, and rewards. MIDCA defines its traces as "a sequence of mental states and mental actions interleaved", with additional formal mathematical notation (Dannenhauer et al. 2018, p. 6).

A sizeable subset of CMAs — including H13, ALMA 2.0 and CARINA — use ontological elements. We distinguish between formal ontologies, axiomatized in a formal language, from informal ones that are described by text; where unspecified, we assume the latter. Formal ontologies are rare. H13 requires a (presumably) formal ontology as input for the employed generalization algorithm (cf. Section 5.3), but this must be supplied externally, as none is specified (Helgason 2013). CARINA (Madera-Doval et al. 2018) encodes certain profiling data (see Section 5.2) via an ontology formulated using the *Web Ontology Language* (OWL). CRAM adopts a hybrid approach: An extension of the *Socio-Physical Model of Activities* (SOMA) — a formal OWL ontology originally designed for modelling physical experiences (Beßler et al. 2021) — standardizes the vocabulary of CRAM's metacognitive experiences (Nolte et al. 2023). Potential sub-symbolic data remain without formal definition; they are envisioned to be linked to the resulting knowledge graph via the KnowRob framework (Beetz et al. 2018) in so-called *Narratively-Enabled Episodic Memories* (NEEMs), as is already the case for physical experiences such as pose transitions undergone by the CRAM powered robot. ACNF also uses an (informal) ontology, the *Cognitve Conceptual Ontology* (CCO) (Crowder et al. 2014, p. 30), which semantically describes cognitrons, the federated building blocks of ACNF's computations. SOMA and CCO overlap in some aspects, e.g., both describe events, states, capabilities, goals and decisions — however, we were unable to compare their respective conceptualizations, as CCO's definitions and potential formalism are, to the best of our knowledge, not publicly available. ISAAC encode its "metareasoning representation" through a semi-formal ontology using a custom knowledge representation scheme that combines graph and frame concepts (Moorman 1997). Meta-AQUA's graph-like traces, which draw heavily from Schank's (1986) *Explanation Pattern* (XP) Theory (Cox and Ram 1999, p. 10), feature some informal-ontology-like commitments (e.g., the relations mentioned in Section 5.2), and have been adopted by MIDCA (Cox et al. 2016). SALS (Morgan 2013) uses Minsky (1975)-style frames with pre-defined slot values.

Instead of *description logics* (the formalism underlying OWL), *active logic* (Miller and Perlis 1993) sees some usage by both MIDCA (Perlis et al. 2013) and ALMA 2.0 (Goldberg 2022) to keep track of inference steps, though their axiomatizations are not detailed in the relevant reports. NARS, like ISAAC, employs its own knowledge representation language, "Narsese", and its traces are sequences of such sentences (Wang et al. 2018).

Finally, some CMAs use looser formalisms. FAtiMA (Mascarenhas et al. 2022) derives its 22 encoded emotions from the OCC model of emotions (Ortony et al. 1988). BICA adopts as a conceptual definition a "Mental State Formalism" (Samsonovich et al. 2009) and a shorter "Schema Formalism" (Samsonovich and De Jong 2005, pp. 2–3) to outline mental-state content.

### 5.5 (B5) Level of Specificity

During our initial literature search, we noticed that CMAs label and interpret traced events (cf. Section 5.2) in three main ways, which we coin *levels of specificity*, each linked to a distinct facet of the traced events:.

*Domain-specific labels* are tied to the problem domain. They describe solutions or issues that do not generalize beyond that domain.

*System-specific labels* refer directly to internal system functions and carry little abstraction.

*Anthropomorphic labels* metaphorically map system functions onto human-cognition terms. Although they borrow everyday vocabulary or from common-sense psychology, authors often give them precise technical definitions that abstract away from domain details.

Placing a CMAs in one category is not always clear-cut because authors sometimes mix levels of specificity and seldom explain their naming choices. The following assignment is based on our subjective classification. As a notable exception, Gordon et al. (2011) argues that anthropomorphic self-models offer three benefits: (1) people naturally anthropomorphize AI agents, so matching human reasoning makes behaviour more predictable and collaboration easier; (2) re-using psychological theory is sufficiently specific while also more efficient than inventing new frameworks; and (3) shared human-like concepts (e.g., memory, attention) promote coherence and interoperability across CMAs. MIDCA's's adoption of Meta-AQUA's anthropomorphic traces (see Section 4.5) provides anecdotal support for this view.

#### ***Domain-specific labels***

Examples of CMAs that use *domain-specific* labels include SCADS (Shrager and Siegler 1998), which operates in the domain of summation using fingers, and Metacat, which focuses on analogy detection. For instance, faced with $abc \rightarrow abd$ ; $xyz \rightarrow ?$, Metacat notices that $a$s and $b$s in $abc$ and $abd$ match and logs an *Identity* observation (Marshall 2006, p. 284), then an event such as *Change letter-category of rightmost letter to successor*, corresponding to the exact insight (Marshall 2006, p. 284). Metacat also uses the label *snag* for failed ideas — an anthropomorphic term — but domain-specific labels dominate.

#### ***System-specific labels***

Soar directly logs the applied operators (Laird 2012, p. 229); KS02 records rules whose conditions were tested or actions executed (Kennedy and Sloman 2002, p. 30). Operators may incidentally be named after anthropomorphic or domain-related concepts. For instance, CARINA mixes readable labels like *new_stimuli_is_detected* with technolect such as *copySMUtoBCPUinput* (Flórez et al. 2018, p. 624). ACT-R seems to also use system-specific labels, e.g., followed strategies including *tree* and *NoCommunication* (Reitter and Lebiere 2012, p. 246).

#### ***Anthropomorphic labels***

*Anthropomorphic labels* are employed, e.g., by SALS (Morgan 2013), MCL (Schmill et al. 2007) and Meta-AQUA (Cox 2011). The latter, for instance, uses terms such as "Memory Retrieval", "Infer", and "Pose Question". As mentioned in Section 5.2, Meta-AQUA links these labels to underlying facts, such as the specific inferences made, which are often domain-specific (Cox 2011); SALS does so, too (Morgan 2013). CRAM names mental events anthropomorphically while its underlying software-component model remains system-specific (Nolte et al. 2023). FAtiMA's (Mascarenhas et al. 2022) emotions come from the The OCC model of emotions (Ortony et al. 1988), also based on human emotions.

#### ***Unlabelled traces***

Two CMA appear label-free: GPME uses metadata fingerprints instead, and D20 stores states and actions as raw tuples (Section 5.2); whether those states/actions are labelled is unclear.

### 5.6 (B6) Episodic Structure

Many CMAs split their memory into distinct time segments but name them differently: *episodes* [ACT-R (Reitter and Lebiere 2012, p. 246) BICA (Samsonovich et al. 2009); MCL (M'Balé and Josyula 2014); Soar (Laird 2012)], *cycle* [KS02 (Kennedy and Sloman 2002)], *phases* [MIDCA (Cox et al. 2016)], *cases* [ACNF (Crowder et al. 2014)], or *events* [CARINA (Caro Piñeres et al. 2019)]. Criteria for where one segment ends and the next begins are often left vague. For instance, MCL opens an episode when it detects an anomaly and closes it once "the anomaly is no longer detected or decays" (M'Balé and Josyula 2014, p. 207). We here briefly describe the other CMAs's approaches for which we have information.

Soar ties each episode to a single processing cycle (Laird 2012, p. 229), which consists of a complete loop through *input* → *operator selection* → *operator application* → *output* (Laird 2012, p. 79). Soar retrieves the most recent matching episode by cueing on a "conjunction of working-memory elements", with options for additional negative features (Laird 2012, p. 231). Afterwards, Soar can retrieve the chronologically adjacent episode, either directly preceding or succeeding the current one (Laird 2012, p. 230).

ACT-R's episode recording is highly task-specific: In a scenario, in which it applies different communication strategies (to later compare their effectiveness), ACT-R records an episode when it receives a "useful fact" via a message sent by another agent (Reitter and Lebiere 2012, p. 246). Usefulness here is identified simply as previously requested information.

The BICA family defines an episode as one mental state or a concurrent set of states, viewed as a graph. It ignores clock time and clusters similar episodes via *cognitive maps*, realized through neural networks for efficient retrieval into working memory (Samsonovich and De Jong 2005; Samsonovich 2006; Samsonovich et al. 2009).

When Metacat proposes a new solution to a domain problem, previous solutions with analogous *themes* (see Section 5.2) are activated in episodic memory (Marshall 2006, pp. 282, 300). Unlike BICA, Metacat does not score similarity via neural networks but quantifies the overlap between the descriptions of previous and proposed solutions (Marshall 2006, p. 300).

MIDCA's object-level runs six phases — *perceive* → *interpret* → *evaluate* → *intend* → *plan* → *act* (Cox et al. 2016, pp. 3713) — each recorded and processed in separate trace segments (see Section 6.3 for details).

## 5.7 (B7) Information-Storage Data Structures

Architectures store their metacognitive experiences in diverse ways; the relevant literature, however, provides limited details, leaving us certain in only a few cases: As mentioned in Section 5.2, some CMAs trace a single, sequential chain of reasoning. Thus, ISAAC (Moorman 1997), MIDCA (Cox et al. 2016) and CARINA (Flórez et al. 2018) utilize lists (specifically, the latter two employ linked lists), though MIDCA later migrated to ordered dictionaries (Mohammad et al. 2020). These might then be accompanied by additional data structures: For example, ISAAC also features a knowledge graph based on frames and slots (Moorman 1997). Soar (Laird 2012, p. 242) records only transitions between episodes rather than the complete episodes, saving space akin to *Delta encoding* commonly employed for other episodic data, such as video streams.

Other architectures encode their metacognitive experiences solely as frames, e.g., SALS (Morgan 2013) and MCL (M'Balé and Josyula 2014; M'Balé 2017), or graphs, e.g., Meta-AQUA (Cox 2011, 2007) and MT21 (Canbaloğlu and Treur 2023). BICA arranges both its mental states and their content as labelled-graph "schemas" with special root nodes representing each schema as a whole (Samsonovich and De Jong, 2003, p. 1031; Samsonovich, 2006, p. 3; Samsonovich and De Jong, 2002, pp. 69–70). Implementation specifics remain sparse.

KS02 represents traces as "a conjunction of [logical] statements that are true for that cycle" (Kennedy and Sloman 2002, p. 12). Similarly, NARS views its experience as a stream of logical statements in its custom knowledge representation language *Narsee* (Wang et al. 2018).

Some CMAs conceptualize and employ custom data structures, including ALMA 2.0, ArmarX, and CRAM: After each step of inference, ALMA 2.0 generates hash-table-based *prospects* for "storing formulas and other information that is identified as potentially satisfying an inference rule" that are similar to logical justifications (Goldberg 2022, p. 136). ArmarX introduces the *ArmarX Interpretable Data Format* (IDF), a recursive compositions of basic data types such as int and string (Peller-Konrad et al. 2023, p. 11). As already mentioned in Section 5.4, CRAM employs so-called *NEEMs*, which combine a knowledge graphs over an OWL-ontology with linked sub-symbolic data such as pose transitions. These are also envisioned for its metacognitive experiences (Nolte et al. 2023).

# 6 (C) Objectives and Evaluation

## 6.1 (C1) Framed Purpose and (C2) Target Issues

We observed that CMAs differ markedly in how they value the examined technology. Soar, for example, does not explicitly model metacognitive experiences; they are simply by-products of episodic memory, which logs every applied operator. Jackson (2019, 2014) notes that TalaMind can in principle record and reflect on past thoughts expressed in near-natural language, yet this capacity appears to have been illustrated only once, in a very brief demonstration. By contrast, for CMAs such as SCADS, MIDCA, Meta-AQUA, and the BICA family, the technique is central to overall operation, and the associated literature describes it in depth.

For CMAs of the latter category, authors claim the technique addresses many challenges on the path to AGI. Research targets include (super)human-level **learning** [ACNF (Crowder et al. 2014); BICA (Samsonovich 2009); CRAM (Nolte et al. 2023); FORR (Epstein and Petrovic 2011; Samsonovich 2010); MCL Family (M'Balé and Josyula 2014; M'Balé 2017); Meta-AQUA (Ram and Cox 1995; Cox 2005a, 2007); PRODIGY (Carbonell et al. 1991); RCAA (Foltyn et al. 2006); REM (Goel and Jones 2011); SALS (Morgan 2010, 2013); SCADS (Shrager and Siegler 1998); Soar (Laird 2012)], **attention** [D20 (Daglarli 2020); H13 (Helgason 2013)], **autonomy** [MCL Family (Schmill et al. 2007); RCAA (Foltyn et al. 2006)] or **emotions** [BICA (Samsonovich 2013b,a); CLARION (Sun et al. 2022; Sun 2018); D20 (Daglarli 2020); H-CogAff (Kennedy 2018); K12 (Kennedy 2012); MAMID (Hudlicka 2005)]. Some systems pursue benchmark performance — variants of the **Turing Test**, for example [K12 (Kennedy 2012); Metacat (York and Swan 2012)] — whereas others aim to emulate **consciousness** [BCP13 (Brody et al. 2013); BICA (Samsonovich and Nadel 2005); D20 (Daglarli 2020); NARS (Wang et al. 2018)] or **self-awareness for human-agent collaboration** CCAs [ACT-R (Birlo and Tapus 2011); BICA (Samsonovich and De Jong 2003; Samsonovich 2009); K10 (Kennedy 2010); MT21 (van Ments and Treur 2021)]. Other purposes were more subtle: **explaining human behaviour** [CLARION (Sun 2018); MAMID (Hudlicka 2005)], **justifying answers** [Meta-AQUA (Cox 2011); Metacat (York and Swan 2012)], **resisting manipulation** [KS02 (Kennedy 2003)], **breaking cyclic behaviour**[22] [Metacat (Marshall 2006); SCADS (Shrager and Siegler 1998); Soar (Laird 2012)], and **recalling prior solution successes and failures** [CRAM (Nolte et al. 2023); H13 (Helgason 2013); Metacat (York and Swan 2012); Soar (Laird 2012)].

Awakening CMAs (Section 4.5) are furthermore motivated by specific shortcomings in their parent systems. ALMA 2.0 improves on its predecessor mainly in allowing nested logical statements (Goldberg 2022). SALS learns to anticipate consequences of mental actions, which its predecessor EM-ONE was not capable of (Morgan 2013). Meta-AQUA augments AQUA's story understanding system (Ram 1989) by selecting compensatory learning strategies when comprehension fails (Ram and Cox 1995). Metacat remedies Copycat tendency to repeat failed solutions (Marshall 2006) and its inability to explain why it prefers one solution over the other (York and Swan 2012).

[22] Kotseruba and Tsotsos (2020, p. 57) already identified this as an important application of the surveyed technique.

Besides Meta-AQUA, CARINA, KS02, MIDCA also employ traces for anomaly detection: CARINA establishes runtime baselines to flag outliers (Madera-Doval et al. 2018, p. 199), KS02 inspects traces to locate rules corrupted by malware [(Kennedy and Sloman 2002); see also Section 6.3], and MIDCA reuses Meta-AQUA's meta-level for the same purposes, plus for switching between fast and slow subsystems (Dannenhauer et al. 2014).

Although CMAs such as Metacat (York and Swan 2012), Soar (Laird 2012, p. 234), Meta-AQUA (Cox and Raja 2011a), and CARINA (Caro Piñeres et al. 2018) name explainability as a primary goal, raw traces are often inscrutable even to experts, necessitating user interfaces or self-explanations when interacting (Cox 2011, pp. 133, 142). Metacat, for example, offers a graphical tool to drill-down into the metadata of individual events (Marshall 2006, p. 287).

Data for BICA family-CMAs remains sparse. When BICA "wakes up" from sleep, it reinstates prior goals and thoughts (Samsonovich and De Jong 2005, p. 4), but more concrete information was not given. The situation is similar for FAtiMA, for which we only know that the CMA, designed to power non-playable characters in video games, can recall emotions and their causes from earlier sessions (Mascarenhas et al. 2022).

Two projects pursue standardisation: MISM surveys shared CMAs features (Caro Piñeres et al. 2014), and CRAM proposes an ontology of metacognitive experiences (Nolte et al. 2023).

Finally, Kennedy (2012) argues that tracing metacognitive experiences is a precursor to ToM: By putting itself in another agent's shoes, a CMA could infer that another agent broke a vase not out of malice, but because it "didn't see the vase or didn't know that it was fragile" (Kennedy 2012, p. 1). The necessary simulation of the other agent's mental state should be learned using the traces of the CMA's own metacognitive experiences as training data.

## 6.2 (C3) Worked Examples

In this section, we consider whether authors exemplify how the respective CMA uses traces of metacognitive experiences. By this we mean, for instance, step-by-step walkthroughs of the operation using tangible domain problems — evaluation scenarios, including case studies, are covered in Section 6 instead.

Concrete examples of trace use are rare, with most relevant reports either omitting them or providing only cursory ones. In the domain problems presented for Metacat and SCADS, for example, episodic memory is treated as a black box (Marshall 2006; York and Swan 2012; Shrager and Siegler 1998). The BICA literature offers just one high-level reference to episodic memory being invoked on waking (Samsonovich and De Jong 2005, p. 4); similarly brief are BCP13's example of a simple rule shutting down processes that run to long (Brody et al. 2013), and H-CogAff's example of how meta-levels may interact (Kennedy 2011). Madera-Doval et al. (2018, p. 625) offered a trace snapshot of CARINA's perception process, and Nolte et al. (2023) exemplarily instantiate CRAM's representational scheme; neither explain how the traces are exploited. M'Balé (2017) offers several small examples of how certain components in GPME process metacognitive experiences, e.g., how they merge recorded episodes or update its response arsenal based on success rates.

Meta-AQUA is the notable exception with well-documented scenarios explained from start to finish — e.g., comprehending stories about a drug bust (Ram and Cox 1995), shooting

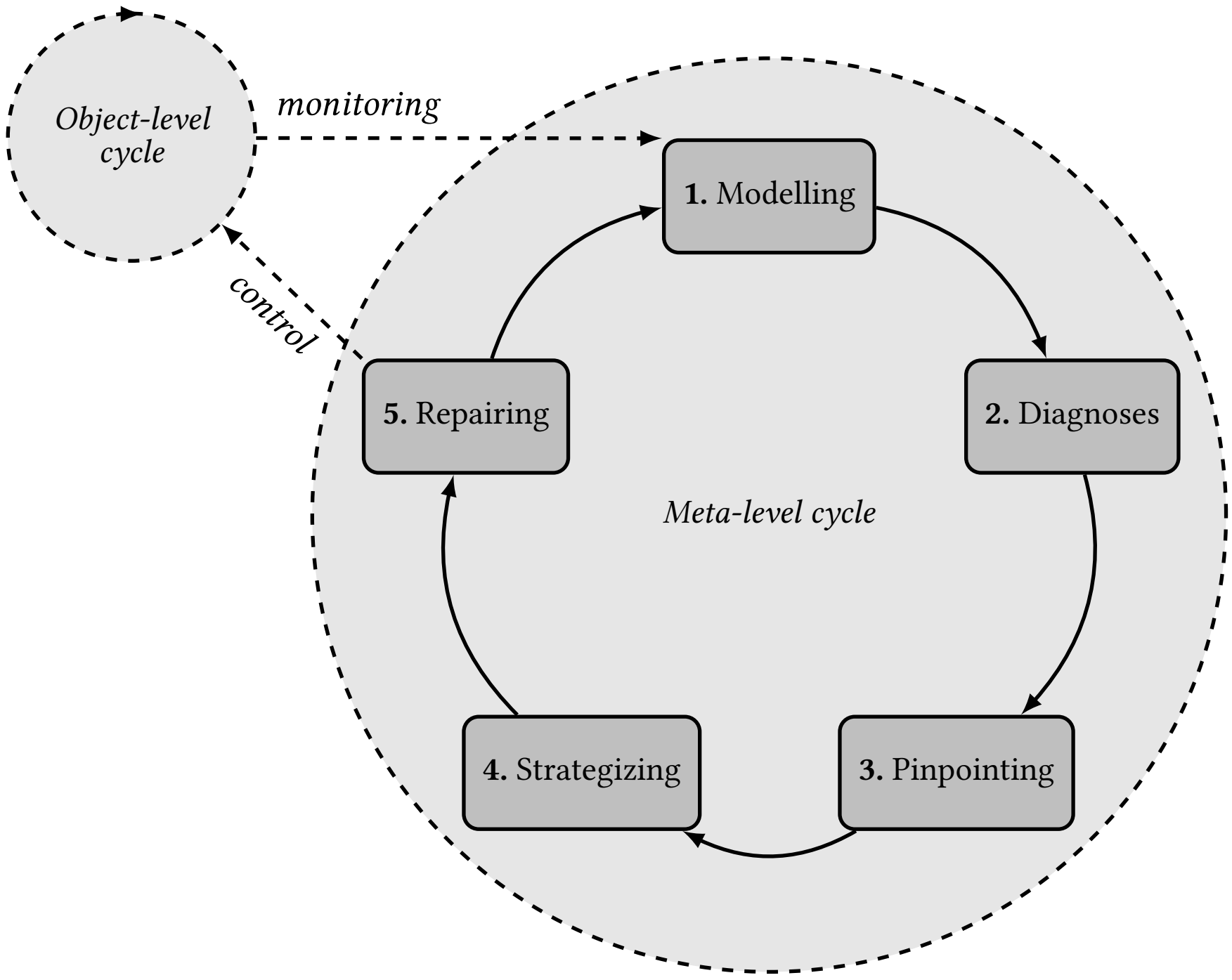

**Fig. 14** Circular flowchart depicting the five phases of a typical metacognitive cycle plus information exchange with the object-level. For CMAs deviating from the Nelson and Narens (1990) model (cf. Figure 10 on page 24), the latter is adapted as necessary

a Wumpus (Cox 2007), or forgetting to fill up gas (Cox 2011) — and explain in detail how traces support understanding. MIDCA, Meta-AQUA's successor, also features a detailed example on it learns consequences of domain-level effects from its trace of metacognitive experiences (Dannenhauer et al. 2018).

## 6.3 (C4) Processing Algorithms

The major cognitive processing of many surveyed CMAs — their *cognitive cycle* — is roughly divided into three subsequent phases: (1) perception, (2) decision, and (3) action. This division appears either explicitly [BICA (Samsonovich 2013a), CARINA (Caro Piñeres et al. 2019), KS02 (Kennedy and Sloman 2002), MIDCA (Cox et al. 2016), Soar (Laird 2019)] or implicitly [Metacat (Marshall 2006), Meta-AQUA[23] (Cox, 1994; Ram and Cox, 1995, p. 216), MISM (Caro Piñeres 2015)]. To process their metacognitive experiences, most CMAs employ strategies analogous to those used in object-level problems, mirroring the object-level cycle phases, resulting in a *metacognitive cycle* [explicitly termed as such by, e.g., MIDCA (Cox et al. 2016)]. This approach offers and facilitates development, particularly for engineers familiar with the object-level. Our qualitative analysis identified five phases within a typical metacognitive cycle (Figure 14), though some CMAs may intertwine or skip phases:

1. *Modelling:* Data collected during monitoring is translated into an introspective model of object-level cognition.
2. *Diagnoses:* Object-level performance is analysed via the constructed model, focusing on potential deficiencies, including errors and sub-optimal behaviour.
3. *Pinpointing:* Causes of identified deficiencies are sought, often using blame-assignment approaches.
4. *Strategizing:* Recovery strategies are devised to overcome identified causes, with one suitable strategy selected.
5. *Repairing:* The selected strategy is detailed and the object-level is directed to execute it.

[23] Although Meta-AQUA does not parse the stories it analyses (Ram and Cox 1995, p. 216) — i.e., it lacks object-level perception — its metacognitive monitoring can be viewed as meta-level perception.

The ongoing monitoring process, conceptualised outside these phases, serves as the meta-level counterpart to sensory perception. Accordingly, the *Modelling* phase corresponds to transforming sensory information into abstract concepts for further processing. The *Diagnosing*, *Pinpointing* and *Strategizing* phases jointly form the equivalent of the object-level decision phase, where the next optimal action is normally sought. Note that meta-level goals seem rarely explicitly represented and can be implicitly summarized as *achieving optimal object-level performance*. Finally, the *Repairing* phase corresponds to the object-level action phase, with control information flow corresponding to object-level manipulation of the ground-level [as conceptualized by Cox and Raja (2011a)].

Since discussing each phase in isolation would require readers to track multiple appearances and algorithms of the same system across scattered sections, we adopt a thematic grouping approach that instead clusters systems by shared techniques or concepts. This perspective facilitates comparison of architectures sharing core ideas, while we may highlight certain architectures featuring noteworthy concepts regarding specific phases.

### 6.3.1 Predefined Pattern-Based Approaches

Some architectures rely on predefined patterns of undesired object-level behaviour during the Diagnosis phase. These patterns can be mapped a priori to error sources and solution strategies, enabling streamlined Pinpointing and Strategizing phase.

Meta-AQUA, for instance, reapplies its explanation techniques from story understanding to traces of its own object-level processing (Cox 1994; Ram and Cox 1995): Different *Introspective Meta-Explanation Pattern*s (IMXPs) associate graph structures called *Explanation of failure* with failure types, pointers to probable failure causes in graph-shaped traces, and compensatory learning strategies. Failure types include contradictions, impasses, surprises, false expectations, unexpected successes, and memory failures (Cox 1995, 1997a). When an *Explanation of failure*, defined as a LISP data structure (Cox 1994), matches part of the trace, Meta-AQUA strategizes a learning goal using the associated learning strategy and data indicated by failure cause pointers. For example, if Meta-AQUA fails to explain a new story and the *Novel Situation Failure* pattern matches the monitored trace, failure pointers indicate insufficient object-level XPs (Meta-AQUA's story domain counterpart to IMXPs for associating narratives with causality). The system then determines learning goals to generalize identified XPs for the misinterpreted situation and invokes appropriate learning algorithms (Cox and Ram 1994). A more illustrative example may be found in (Ram and Cox 1995, pp. 216–218). Note that IMXPs are not only defined for failures but also for correct behaviour, e.g., to identify when the system managed to make a successful prediction.

AML11, strongly inspired by Meta-AQUA, adapts error explanations specifically for CBR systems. When the CBR-suggested solution deviates from the ultimately found solution, learning mechanisms trigger similarly to Meta-AQUA, but based on solution differences rather than processing trace pattern matching.

Metacat searches for "patterns in its own behaviour in much the same way that Copycat [Metacat's parent, cf. Section 4.5] perceives patterns in letter-strings" (Marshall 2006, p. 284). Although details remain unclear, Metacat monitors object-level event traces for undesirable patterns, such as repeated proposals of similar but incorrect ideas. Upon detection, Metacat *clamps* the pattern, reducing the likelihood that defining themes and concepts will be considered by its probabilistic codelets, thus allowing exploration of alternative ideas. When no important events occur for extended periods, Metacat can unclamp patterns to mix newly generated ideas with suppressed topics. Conversely, if excessive clamping occurs without progress, Metacat will "gracefully" quit the solution search (Marshall 2006, p. 289).

Kennedy (2018, p. 7) planned to use predefined trace patterns to detect "involuntary" motivational state changes in H-CogAff; no follow-up work was found.

Foltyn et al. (2006) provide details about RCAA's monitoring process: They conceptualize listeners registered to its working memory that asynchronously update episodic memory, which is further processed into *detailed* and *aggregated* models. The latter contains statistics about the former, such as event frequency extracted from the explicit event history. These values, including "agent's success ratio [. . . ] or the current system load", are weighted separately for each RCAA component to compute an overall utility score for currently employed algorithms (Foltyn et al. 2006, p. 4). Low-scoring algorithms may be discarded or restructured, for which *inductive logic programming*-based planners also consider monitored object-level events. This approach is predefined in that considered values and component-specific weights seem set a priori.

Brody et al. (2013) also focus entirely on monitoring and modelling. Their approach involves an inference engine for *active logic* — a temporal logic linking step-wise inferences to incrementing time symbols — that adds reserved predicates *Applying*$(t, r, a, c)$ to the database whenever "at time $t$, the inference rule [. . . ] $r$ was applied with the antecedents [. . . ] $a$ to derive the consequent [. . . ] $c$" (Brody et al. 2013, p. 6). These facts become available to all inference rules for application in step $t + 1$, thus becoming subject to reasoning themselves. Notably, this approach is intended as a riposte to metacognition that only analyses completed reasoning processes retrospectively; instead, it enables fine-grained reflection of individual inferences while the reasoning process is still underway.

The reasoning system ALMA 2.0 supports predefined nested logical rules, applied to common reasoning examples such as reasoning with defaults and reasoning over past beliefs by quoting them in nested rules (Goldberg 2022).

### 6.3.2 Machine Learning-Based Approaches

Instead of using rigid, predefined patterns during the Diagnosing phase, many CMAs employ machine-learning techniques that detect deficiencies more flexibly. We distinguish three approaches:

### *Traning-based Anomaly Detection*

Some meta-levels learn normal object-level behaviour during an initial *training phase* [some authors use different terms such as "Self-Familiarization" (Kennedy 2011, p. 236)], with any nonconforming activity considered anomalous during the subsequent *operational phase*.

The earliest such approach is KS02 (Kennedy and Sloman 2002). Inspired by *signature-based intrusion detection systems* [see (Gupta et al. 2023)], KS02 builds a "database of 'normal' patterns" expected every cognitive cycle trace (Kennedy and Sloman 2002, p. 13). KS02 generates *detectors* for each traced event, weighting them proportionally to their occurrence frequency across cycles. During operation, anomalies are identified when detectors with 100% weighting fail to match events in any cycle. KS02 then attempts to identify damaged rules by absent events (rules that failed to fire) and repair via backup mechanisms. Kennedy and Sloman (2003) extended KS02 to count "unfamiliar activity" as anomalies but only react when accompanied by performance decreases, identified through simple thresholds (e.g., when energy levels fall below minimum training values). Unlike Meta-AQUA's hard-coded graph queries (Section 6.3.1), KS02's trained detectors correspond to simpler atomic queries.

Notably, Kennedy (2003, p. 118) considers protecting monitoring components essential: They should remain immutable to prevent meta-level manipulation — even the object-level must not tamper with the trace production. KS02 threat protection complicates monitoring: KS02 possesses two object-levels, each with meta-levels that also monitor each other (recall Section 4.4). The meta-levels constantly interleave phases asynchronously, monitoring not only the other meta-level's operational phases but also training phases and transitions (Kennedy and Sloman 2002, pp. 15–19). A training meta-level may even corrupt another operational meta-level's object-level to monitor the exercised control. Kennedy and Sloman (2002, pp. 18) note a limitation in their setup with two equal meta-levels: a meta-level with anomalous behaviour is blind to repair interventions by the other meta-level, requiring a third equal meta-level. Kennedy (2010, 2011, 2012) envisions similar pattern learning for K10s, H-CogAff, and K12s, respectively.

### *Continuous Learning Approaches*

The second approach foregoes training/operation division, with meta-levels constantly manipulating object-levels through experimentation to find improvements.

SCADS's meta-level varies symbolically represented object-level algorithms in 5% of runs by rearranging steps (Shrager and Siegler 1998). Heuristics analyze previous run traces to verify efficiency and detect/delete redundant calculation sequences, with consistently more efficient variations being added to the repertoire.

Morgan (2013, p. 81) conceptualizes SALS's monitoring phase as event reification by transforming raw knowledge base changes (at any layer; Section 4.4) into symbolic events processable by higher layers. Each atomic change in lower-layer knowledge bases is communicated asynchronously to the monitoring layer via buffering streams, then recorded in layer-specific *Partial State Event Knowledge Bases.* A second stream of *activated resources* (physical or mental actions executed by lower layers) triggers learning components. SALS improves planning by computing *transframes* (Minsky 1975) between before/after states of activated resources, looked up from the Partial State Event Knowledge Base. During planning, the generated transframes function as consequence hypothesizes of lower-layer activated resources.

H13, FORR and ACT-R all use their episodic memory to identify processes or strategies that are useful to achieve their goals. H13 (Helgason 2013) heuristically computes utility scores for past processes associated with goals or goal patterns (cf. Section 5.3) based on another heuristic, goal priority, back-propagated through process history. When new goals match episodic memory patterns, processes deemed useful via utility value are activated to contribute towards news goal (Helgason 2013, p. 134–138). When processes successfully contribute to goal achievement, H13 reinforces their utility and durability scores. H13 may also delete pattern-process associations from episodic memory when decaying durability values fall below some threshold. ACT-R (Reitter and Lebiere 2012, p. 246), following *Instance-based learning*, constructs episodic memories of its object-level actions, associating them with employed strategies and successfulness. Episodes involving identical strategies merge into a "blended episode", representing average strategy success. FORR also weighs object-level heuristics suggesting specific actions based on training instances extracted "from traces of its [FORR's] problem-solving behaviour" (Samsonovich 2010, p. 221). Beyond action correctness, instance associate with "problem state, the available choices there, and the decision made" (Epstein and Petrovic 2011, p. 51).

A combination of reification and Hebbian learning is used for MT21 intended to describe the use of mental models (Canbaloğlu and Treur 2023; van Ments and Treur 2021). Mental models in this approach are networks with nodes having activation values and connection strengths that change over time using the "fire together, wire together" heuristic of Hebbian learning. MT21's specification leaves the semantic content of object-level network nodes unconstrained, and nodes need not be semantically opaque as in most neural networks; for example, there can be nodes representing particular stimulus conditions, nodes for specific reactions, etc. The various update coefficients of nodes in lower-level networks are reified as nodes in higher-order networks, which then model the reactivity and plasticity of lower-level networks and thus control their adaptability. This approach also eliminates separation between training and operational phase — network states change continuously based on input signals and network response parameters, with network connectivity itself also being changeable.

We are not entirely clear whether D20's (Daglarli 2020) approach fits into the first or second category, i.e. whether it is offline or online. As already mentioned in Section 5.2 D20 employs a variant of Q-learning, in which the reinforcement values are not imposed from the outside after each action, but are determined by D20's meta-level itself, again implemented as a combination of neural networks (Daglarli 2020).

#### ***Case-Based and Explanation-Based Reasoning***

As noted in Section 2.1.3, some architectures use techniques reminiscent of CBR:

PRODIGY (Carbonell et al. 1991) develops search trees on the object-level to solve domain problems, with nodes representing world states. Starting from the root (the current state), the tree expands iteratively until reaching a node representing the goal state. This involves selecting promising world states to expand and operators to apply, creating child nodes with updated world states. So-called *control rules* advise which nodes to select, reject or prefer. PRODIGY's meta-level employs different optimization strategies. One module develops new control rules: candidates are included if their expected search time reduction, estimated from past problem solution traces, outweighs the overhead of permanently checking rule applicability. Although PRODIGY's rejecting rules suppress certain ideas similarly to Metacat's pattern clamping, they facilitate long-term learned behaviours rather than short-term reactions. Another CBR-like module guides object-level search by remembering past problem solutions.

Three further CMAs use CBR, though relevant reports lack details: Metacat stores previous problem solutions and attempts to reapply them to novel but similar problems (Marshall 2006, pp. 303–304). ACNF uses CBR to learn new action plans (Crowder et al. 2014). CARINA employs CBR as an episodic memory (Caro Piñeres et al. 2019, p. 80).

As mentioned in Section 6.3.1, AML11 also incorporates a CBR system; however, it functions as the monitored and controlled object-level rather than the meta-level.

### 6.3.3 Hybrid Approaches

CARINA profiles its cognitive functions to learn their expected behaviour (Caro Piñeres et al. 2018, p. 532), representing expectation as logical axioms (Caro Piñeres et al. 2019, 85). Sensor information violating these constraints triggers a *violation of expectation* (Caro Piñeres et al. 2019, 85). CARINA then attempts to explain the anomaly either by checking whether past explanations apply (similar to CBR; see Section 2.1.3) or by analysing reasoning traces. The only additional information provided is that "the reasoning trace analysis strategy performs a more complex reasoning" (Caro Piñeres et al. 2019, 85) via querying. This lack of detail is problematic since only reasoning trace analysis is relevant to this review, as the learned expectations represent aggregated, non-episodic data.

MIDCA employs a hybrid approach, where only the top-down *K-track* uses metacognitive experience traces similarly to Meta-AQUA when expectation violations are detected. The complementary *D-track* uses machine learning models (e.g., *Growing Neural Gas* (GNG) networks and Bayesian networks of ontologies) to detect anomalies by comparing current observations with statistically learned baselines (Cox 2013; Cox et al. 2013). Only when the *D-Track* identifies anomalies does MIDCA investigate further via the computationally intensive *K-Track*. Recall from Section 5.6 that MIDCA's object-level and meta-level each comprise six phases: *perceive* → *interpret* → *evaluate* → *intend* → *plan* → *act* (Cox et al. 2016, p. 3713). Accordingly, MIDCA's meta-level first perceives a sub-trace of the complete monitored object-level trace — a fixed-length suffix representing the last object-level phase (Cox et al. 2016, p. 3716). MIDCA then inspects the trace to identify expectation violations and, if found, generates corrective goals, selects which to pursue, plans, and executes necessary actions (Cox et al. 2016, p. 3716). Note that metacognition does not run parallel to object-level cognition in this conceptualization.

### 6.3.4 Unclear Approaches

Laird (2012) mentions that Soar's episodic memory offers multiple capabilities. For instance, it enables countering repetitions by allowing Soar to "move forward through episodes to determine which action(s) it took in similar situations, as to avoid repeating them in the present situation" (Laird 2012, p. 233). Regarding other capabilities such as learning consequences of its own actions (represented as *operators*), uncertainty remains whether these associate with metacognitive experiences, as given examples focus on domain-level actions. Theoretically, these might transfer to object-level cognition, which is also partly represented as operators (such as episodic memory retrieval itself). Soar must deliberately

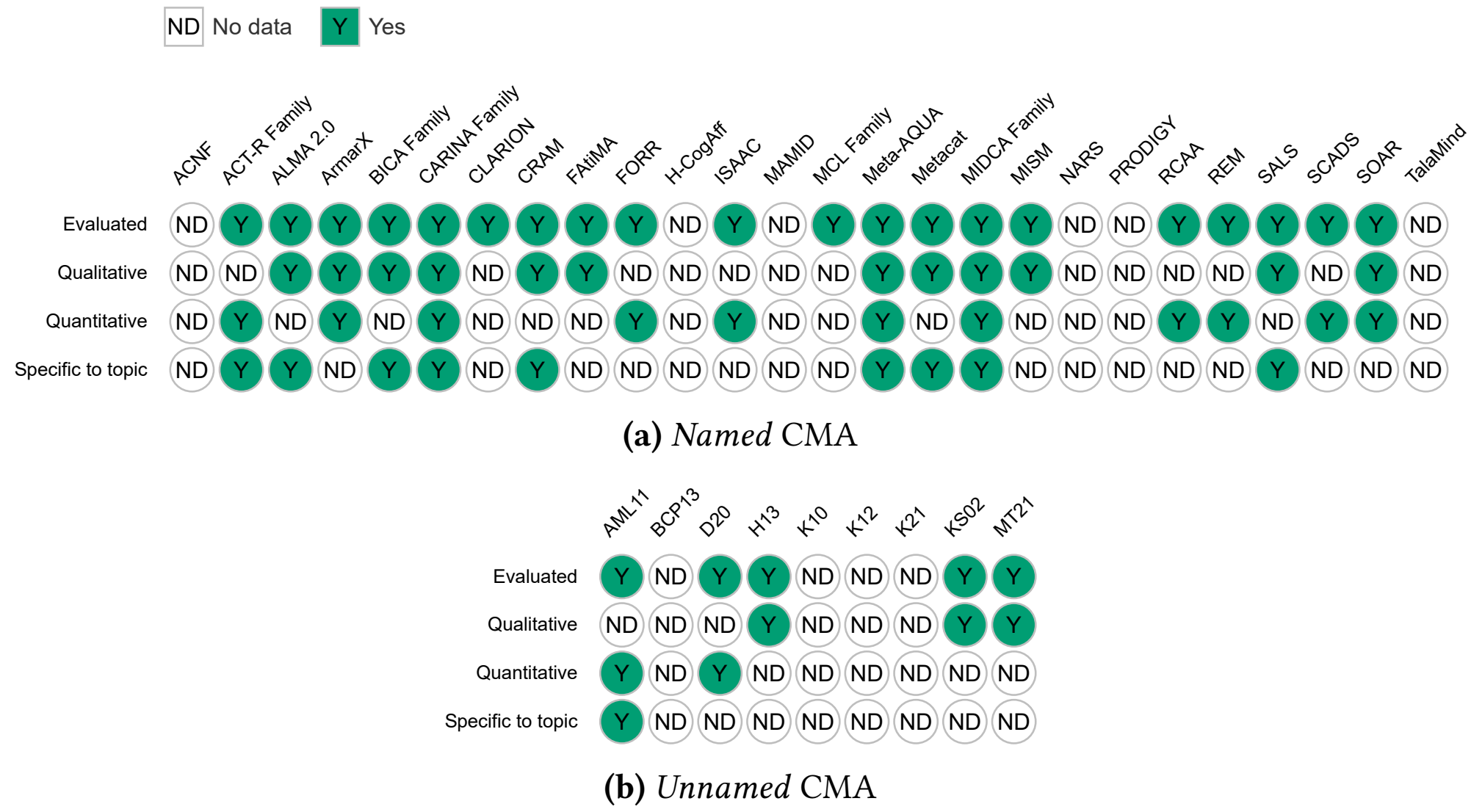

**(a)** *Named* CMA

**(b)** *Unnamed* CMA

**Fig. 15** Whether each CMA has been evaluated in some relevant report, whether any evaluation was qualitative or quantitative, and whether they were specific to performance improvements from using metacognitive experiences. Y = yes; ND = no data

query its episodic memory; however, when it does so and how it uses retrieved information appears implementation-specific for each Soar agent.

Numerous CMAs (BICA, CLARION, ISAAC, and others) lack sufficient detail in relevant reports for us to be confident about understanding their algorithms. For instance, BICA's authors conceptualize belief revision for failed expectations (Samsonovich 2006, p. 6) and mental task distribution across different mental states (Samsonovich et al. 2009). In one example, BICA's mental state *I-Meta* controls imagined mental states *I-Imagined* for planning, making envisioned movement steps closer to the goal more likely be spun further (Kalish et al. 2010). However, the latest relevant report states that "Details of operation of the metacognitive component should be developed based on empirical studies" (Samsonovich 2013b, p. 103).

## 6.4 (C5) Test Scenarios and (C6) Evaluation

### *Evaluation Challenges for Metacognitive Experiences*

In most relevant reports, metacognitive skills allow CMAs to solve new problems instead of improving the solution of already solvable problems (cf. Section 6.1). While many CMAs may be well-evaluated, the specific technique of leveraging metacognitive experiences is often assessed qualitative as proof of concept rather than through quantitative comparative testing; exact data for each CMA appears in Figure 15. Typical ablation tests comparing systems with metacognition activated versus deactivated are often impractical, as the latter might fail entirely when lacking graceful degradation.

SCADS (Shrager and Siegler 1998) illustrates this problem, with its purpose being not efficient summation but mimicking how children learn summation strategies (cf. Section 4.4).

Accordingly, Shrager and Siegler (1998) evaluated whether SCADS matches observations from child behaviour studies, such as detecting identical strategies and tending towards simpler strategies. That SCADS matches children on numerous phenomena is taken as support for claims that the CMA provides an abstracted, explanatory model of human cognition. However, isolating the effect of SCADS' metacognitive experiences is challenging since they are crucial for its learning capability (cf. Section 6.3). Removing these without substitute techniques would likely eliminate the system's learning ability, rendering it non-functional for its intended purpose. Therefore, comparing SCADS with and without trace analysis is impractical — comparing functional vs non-functional systems yields no meaningful insights.

This highlights a significant challenge: without viable methods to compare metacognitive experience processing against other techniques, ascertaining their unique performance contribution is difficult. One potential approach involves reframing system purposes for evaluation. For instance, if we consider SCADS' purpose as efficient summation rather than simulating children's learning, we could compare its object-level alone (capable of summation but not learning) with the complete system. However, this is retrospectively unfeasible with the information provided by Shrager and Siegler (1998), who only discuss the complete system's performance before and after learning. Certainly, we could assume SCADS' object-level performance equals the complete system's pre-learning performance. However, as described in Section 6.3, SCADS' meta-level introduces algorithmic variations in $5\%$ of cases to discover more efficient calculation sequences. Many random variations are likely inefficient or produce incorrect results. Conversely, without the meta-level, the object-level would consistently use correct, though not optimal, known algorithms without such random errors. This suggests that object-level performance alone might exceed the complete system's pre-learning performance. Consequently, evaluations not specifically focused on processing metacognitive experiences are challenging to transfer to that context.

#### ***Quantitative Evaluation Results***

Only a small minority ($\approx 17\%$) of examined CMAs had relevant reports containing a quantitative ablation test where identical agents performed tasks both with and without the surveyed technology enabled. Specifically, we found such evaluations only for approaches employed by ACT-R (Reitter and Lebiere 2012), AML11 (Arcos et al. 2011), BICA (Kalish et al. 2010), CARINA (Caro Piñeres et al. 2013), Meta-AQUA (Cox 1996, 1997b), and MIDCA (Dannenhauer et al. 2018; Cox et al. 2021). The latter five, each running as a complete system with fully utilized metacognitive experiences, consistently outperformed more naïve versions. For instance, in Caro Piñeres et al.'s (2013) experiments, the CARINA-based FUNPRO tutoring system was tested across multiple sessions with progressively fewer resources available, measuring successful resource recommendation rates. Comparing system performance with and without memory event tracing activated revealed that metamemory-driven adaptation consistently yielded higher proportions of valid resource retrievals (e.g., $98\%$ vs. $71\%$). In contrast, activating instance-based learning (cf. Section 6.3.2) generally impairs ACT-R's performance on specific problem solving tasks, except when facing randomized or downsized problem variations, where it also outperforms naïve versions (Reitter and Lebiere 2012).

### *Alternative Evaluation Approaches*

Some CMAs where evaluated on measures other than task performance. Similar to SCADS, CLARION focuses evaluation on comparing agent performance to human performance on expertise-requiring tasks, particularly whether artificial agents make similar errors, such as overconfidence tendencies in wrong answers on certain tasks (Sun et al. 2005). Developers of both Metacat (Marshall 2006) and ALMA 2.0 (Goldberg 2022) discuss additional skills compared to predecessors, exemplified through case studies. Madera-Doval et al. (2018) present another case study where CARINA's meta-level exemplarily identifies an anomaly based on learned expected runtime of object-level cognitive functions. MISM as a metamodel and CRAM's scheme SOMA were evaluated using methodologies strictly defined in their respective development processes: MISM through comparison with CMAs not contributing to its initial design (Caro Piñeres et al. 2014), and CRAM through checking whether the representational scope defined via *competency questions* is met (Nolte et al. 2023) — a common ontology evaluation technique from Grüninger and Fox (1995). Morgan (2013, pp. 139–151) analyses computational complexity of metacognitive monitoring and control in SALS.

### *Indirect Evaluations and Domain Limitations*

Numerous reported evaluations did not specifically target the technique of interest — their success may only indirectly support leveraging metacognitive experiences. Several studies, including case studies, were performed for D20 (Daglarli 2020), FAtiMA (Mascarenhas et al. 2022), RCAA (Foltyn et al. 2006), REM (Goel and Jones 2011), and MIDCA (Dannenhauer et al. 2014; Cox et al. 2016; Mohammad et al. 2020; Cox et al. 2021). ISAAC was tested by answering identical reading comprehension questions as human participants for short science fiction stories, with answers graded anonymously by teachers alongside the human responses (Moorman 1997). Statistical tests showed ISAAC's performance to be comparable to or better than human readers.

Importantly, most experimental evaluations involve very limited domains: **simulated games or environments** [ACT-R (Reitter and Lebiere 2012); BICA (Kalish et al. 2010); FAtiMA (Mascarenhas et al. 2022); FORR (Epstein and Petrovic 2011); MIDCA (Dannenhauer et al. 2014; Cox et al. 2016; Mohammad et al. 2020; Cox et al. 2021); RCAA (Foltyn et al. 2006); REM (Goel and Jones 2011)], **analogy puzzles** [Metacat (Marshall 2006); MCL (M'Balé 2017)], **simple communication** [CARINA (Caro Piñeres et al. 2013); MT21 (Canbaloğlu et al. 2023)], **arithmetic** [SCADS (Shrager and Siegler 1998)], and **story comprehension** [CLARION (Sun et al. 2006); ISAAC (Moorman 1997); Meta-AQUA (Cox 1996)]. MIDCA, for example, cares for plants on $10 \times 10$-tile boards (Cox et al. 2021) or controls a single light switch (Dannenhauer et al. 2014). We identified only one experiment targeting the real world, involving D20 controlling a robot (Daglarli 2020) — however, even that was limited to tic-tac-toe level games.

## 7 Discussion

This paper set out to address the question: *How do models of cognition model their own cognition?* To enable a systematic review, we focused the scope on Computational Metacognitive Architectures (CMAs) that model, store, remember and leverage metacognitive experiences — introspective interpretations of their own cognitive processes. We conducted a systematic search across 16 research databases and identified 101 relevant reports covering 35 separate CMAs. Using 20 different data items, we collected information across three dimensions: model context, data content and structure, and objectives and evaluation.

While the examined approaches are highly diverse — ranging from symbolic logic-driven systems to hybrid and fully emergent frameworks — common themes emerged that highlight both the promise of metacognition as well as persistent limitations in the field. In the remainder of this chapter, we discuss results presented in Sections 3 to 6, as well as limitations of our work and how it may inform future research directions.

## 7.1 Summary and Interpretation of Results

We begin by summarizing our direct findings regarding the posed research question. The most fundamental observation is that most CMAs follow models of human metacognition established in psychological research (Section 4.4). For example, most architectures maintain a clear distinction between object-level and meta-level cognition, following Nelson and Narens (1990). Variants of this principle introduce additional levels for meta-meta cognition, either equal in status or hierarchically arranged. Only a few exceptions process cognition and meta-cognition through a unified component.

Typically, traces of episodic metacognitive experiences are modelled as event-centric (Section 5.2). The chain of events is understood either as linear sequences or, to model parallelism and causality, as complex graph structures (Section 5.2). Events are usually categorized symbolically (Section 5.1), using system-specific or domain-specific labels, or terms borrowed from (household) psychology (Section 5.5). They often serve as anchors for numerous symbolic and sub-symbolic metadata, including process inputs and outputs, justifications, confidence values, and metrics for indexing and similarity-based retrieval (Section 5.2). The type and quantity of metadata vary considerably depending on each CMA's algorithmic requirements. Some CMAs actively reduce data collection to accommodate database limitations by combining several calculation steps into more abstract events or including only the most important events (Section 5.3). At least $40\%$ of metacognitive experience models are formally or semi-formally specified, mostly based on logical systems and ontologies, some of which are axiomatized and formally defined (Section 5.4).

The key role of these traces is retaining sufficient context about the agent "thoughts" to enable deeper introspection, identifying faulty inferences, and trigger appropriate remedies (Section 6.1). Various approaches fall under this description: examining knowledge that led to incorrect conclusions to adjust object-level models, breaking out of looping behaviour, or using prior inference to optimize computation times. The investigated CMAs employ diverse methods and algorithms, from pre-defined pattern matching against the trace to autonomous sub-symbolic machine learning methods with or without a dedicated training phases, and CBR-inspired strategies that explicitly use past experience (Section 6.3). To enhance explainability, some CMAs also produce user-facing "reasons" for their solutions, referencing the recorded chain of insights.

Given the number and diversity of identified approaches, memories of metacognitive experiences are clearly a widely recognized method employed by CMAs. Since emerging in the 1990s, a steady stream of new CMAs have adopted such strategies (Section 4.1), including the prominent models ACT-R, CLARION, and Soar. Despite their prevalence, researchers place varying emphasis on this technique in relevant reports. Some focus heavily on it, while others only mention it briefly. As reported in Section 3.2, some data items are explained in great detail, while others lack sufficient information. Importantly, the research community has yet to agree on a unified terminology (Section 4.2). We observe that the word “trace” occurs particularly frequently, but even that appears in three different semantic contexts. The terminology developed in this review — independently and prior to the terminological analysis, referring mostly to “traces” or “episodic memories” of “metacognitive experiences” — aligns with psychological theory, but is not necessarily better than the variety of terms currently in circulation; still we put this forth as a starting point for a unified language as with explicit definitions.

Further evidence strengthens the impression that the collective scientific reporting in this field is fragmented and lacks transparency: only a minority of architectures make source code publicly available (Section 4.3) or fully disclose the underlying representation schemes (Section 5.4), data structures (Section 5.7), and algorithms (Section 6.3). For example, while structuring metacognitive experiences in episodes (Section 5.6) and filtering events by importance are common practice (Section 5.3), how episodes boundaries are defined and event importance is determined remains largely unclear. Such gaps likely hinder progress, as scientists must reinvent methods while struggling to compare new approaches with existing ones as long as those remain vague.

One contributing factor may be the fragmented research community, where a single well-connected network produces over $50\%$ of relevant reports while other scientists typically publish individual papers on their CMAs (Section 3.3). Furthermore, excluding conceptual inspiration (which we did not test here), and successive versions of the same CMA families, only two CMAs explicitly build upon previous CMAs (Section 4.5).

This fragmentation is also evident in evaluation practices (Section 6.4): regarding episodic metacognitive experiences, relevant CMAs were at most quantitatively compared with naïve variants of themselves, never with each other. This limits evaluation informativeness, preventing determination of which approaches are most successful or merit further research. Additionally, all CMAs were tested in highly restrictive domains, essentially “toy problems”. While this aligns with the criticism of Kotseruba and Tsotsos (2020) regarding the evaluation of cognitive architectures in general (i.e., not limited to metacognition), it is surprising given that numerous authors cite autonomy, human-level intelligence and explainability as clear objectives (Section 6.1) — characteristics particularly important when CMAs meet the real world embodied as robots, and especially in *Human-Robot-Interaction* (HRI). One might expect such domains to receive more attention than the single relevant study, which was also limited to fairly simple games. Consequently, we characterize the initial evidence for the effectiveness of metacognitive experiences in CMAs as cautiously promising.

Overall, our observations support the general impression that the field is fragmented (Brown 1987; Cox and Raja 2011a; Caro Piñeres et al. 2014), and inadequately evaluated (Kotseruba and Tsotsos 2020). In the Introduction, we cite Gordon et al. (2011, p. 296), who argue that research focuses on algorithms rather than representation, leaving models of mental representations under-researched. Our qualitative and quantitative data — specifically the limited algorithmic detail in relevant papers — suggest the problem lies more in focusing on conceptualizing full-featured architectures. While this is undoubtedly important, research papers must provide more detailed information on each component under consideration to ensure reproducibility and comparability.

## 7.2 Limitations

Despite its comprehensiveness, this review has limitations that must be considered when interpreting results.

### 7.2.1 Limitations of the Evidence

As noted in Section 3.2, data availability and detail vary considerably across items and CMAs. This imbalance may introduce bias in corresponding chapters, as we can only report information we were able to collect. While reporting on items with limited data is necessarily restricted to few CMAs, chapters with abundant detailed data will be more representative. Furthermore, we cannot guarantee the collected data's reliability, although we avoided recording obviously incorrect information[24].

Most identified CMAs rely on symbolic or hybrid data (Section 5.1). Sub-symbolic-only approaches are notably rare: we identified only D20 (Daglarli 2020), with detailed information for just 11 of our 20 items. This could indicate under-representation of fully end-to-end, emergent metacognitive architectures — those whose cognitive capacities grow solely from training neural networks on raw data rather than relying on predefined knowledge bases and algorithms — in our synthesis. However, evidence suggests this imbalance may reflect the current research landscape accurately: Kotseruba and Tsotsos' (2020) comprehensive overview identified no emergent architecture as metacognitive. Notably, their criteria for a "metacognitive" architectures were more generous than our requirements for episodic metacognitive experiences, yet no emergent systems qualified. We see two plausible explanations: first, emergent architectures form a relatively small minority of architectures overall, at least among those analysed by Kotseruba and Tsotsos, p. 25. Second, emergent architectures tend to be less transparent, as understanding how their sub-symbolic processes arrive at specific outcomes remains notoriously difficult even with full access to the underlying networks (Kotseruba and Tsotsos 2020, p. 24). This raises non-trivial problems for implementing metacognitive principles: if the underlying processes are poorly understood, how can systems be designed to reflect them? How could we verify whether a network truly leverages self-information metacognitively if even its object-level reasoning remains a black box? While these findings derive primarily from data collected in 2018 (Kotseruba and Tsotsos 2020) and our synthesis spans publications only through 2023, very recent work beyond these timeframes could introduce new developments in emergent metacognitive architectures. Nevertheless, within the current body of evidence, emergent approaches incorporating metacognitive functionality, including metacognitive experiences, remain exceptionally rare.

### 7.2.2 Limitations of the Review Process

Our paper search relied on specific search terms ("metacognition", "monitoring", "tracing", "cognitive agent", etc.) following an iterative but unsystematic initial literature exploration. While the query was designed to be broad and inclusive, it may have missed projects that frame their metacognitive components using alternative terminology as observed in Section 4.2 (e.g., "self-reflection", "introspection", "log analysis", "autobiographic"). Future research could address this issue by conducting a second iteration of this review based on the systematic terminology analysis in Section 4.2.

[24] An example is item (A6) and Thórisson and Helgason's (2012, p. 10) claim that ACT-R is "a predecessor to Soar" — inaccurate given that Soar was described by Laird and Newell in 1983, while ACT-R was first presented by Anderson in 1993, although they influenced each other over the years.

Our eligibility criteria emphasised explicit references to cognitive architectures and metacognition. This focus may favour well-established or more academically oriented frameworks, while overlooking newer or industry-based approaches not positioned under the cognitive architecture banner. Furthermore, our Definition 3 of metacognition is strict and excludes systems that observe object-level actions rather than object-level cognition. For instance, MCL3 (M'Balé and Josyula 2013) traces only "primitive" actions that the host agent performs in the environment, not its internal events, and was therefore excluded. We identify related approaches in Section 2.1.3 to help mitigate these limitations.

In the screening process (Section 2.4), each paper was evaluated by one screener except in ambiguous cases, and data items were collected by only one screener for most relevant reports, i.e., we mostly forego duplicate screening. While this approach kept the screening process feasible within the project's time constraints, it may introduce subjectivity and inconsistency in categorizing records. Similarly, our grouping of CMAs into "families" involves interpretation that, while carefully done, could inadvertently overlook minor but possibly meaningful architectural differences.

## 7.3 Outlook

The challenges identified in this review call for a more unified yet competitive research landscape. We believe that an agreed-upon terminology, open-source code, precise model specifications, and explicit algorithm details will promote exchange of ideas between cognitive architectures. First, successful components could be better adopted, and accumulated metacognitive experience could be shared, mutually providing training data for emergent machine learning approaches — both would advance the field as a whole. Second, improving reproducibility and facilitating comparisons and benchmarking will help identify the most promising research directions.

This also involves testing systems on more complex, dynamic domains that go beyond toy problems, targeting the real world as the ultimate testing ground. In experiments unrelated to metacognitive experiences, several CMAs investigated in this review (other than D20) have operated in real-world settings when embodying and controlling robots; some, such as ArmarX and CRAM, are specifically tailored for this purpose. A first quantitative comparison seems within reach. Meanwhile, research must create alternative evaluation standards: while common ablation tests can isolate the contribution of metacognitive experiences to overall performance, this is less obvious when contrasting distinct systems. We argue that this requires test platforms targeting metacognitive skills such as self-correction and goal-driven learning, for example as part of the *RoboCup* (Asada and Kitano 1999; Wisspeintner et al. 2009).

Such standardization and rigorous testing are critical, given that cognitive science often uses cognitive architectures to test theories of human cognition. Equally, both the AI community and the society in general have much to gain from employing metacognitive experiences. Rapid change is underway, driven by deep learning breakthroughs: neural networks increasingly enter everyday life — from household robots that mop and vacuum (Lee 2021; Huang 2023), self-driving cars (Kim and Joe 2022), and virtual assistants (Chamola et al. 2023) to systems designed to replace human workforce (Tschang and Almirall 2021). However, particularly in critical areas such as medicine or traffic, testing and ensuring the reliability of AI systems is essential. Memories of metacognitive experiences in AI systems not only enable introspective insights by the systems themselves but also make their processes transparent and comprehensible for users, e.g., through interfaces such as Metacat's (Section 6.1). Unlike symbolic or hybrid systems, where interpreting cognitive processes is straightforward (which might be why most relevant reports do not address interpretation and monitoring, implicitly taking them for granted), emergent systems demand a strong research focus towards interpretation. Although not framed as metacognition, such interpretation is what *Explainable Artificial Intelligence* (XAI) is largely aimed at (Gunning et al. 2019). The European Union in particular drives forward research programs and legislation promoting XAI, such as the *Horizon Europe Program* (Eurpoean Commission 2024) and the *AI Act* (Pavlidis 2024).

Current explanation methods have been observed as being inaccurate themselves or even misleading (Rudin 2019). Although painting a complete picture is beyond this study's scope, we conclude by discussing a current topic highly relevant to this systematic review. With the surge in LLM research, investigating whether (meta)cognitive capabilities may be useful to these systems or already emergent has become fruitful. However, because the field is new and the systems involved are complex and semantically opaque, no consensus exists about the presence and qualities of metacognitive abilities in LLMs. For example, metacognition in LLMs has been claimed, such as GPT4 appearing able to judge the skill level needed to solve mathematical problems (Didolkar et al. 2024), while other results dispute metacognition in LLMs, including studies showing that LLMs are poor judges of confidence in their own answers (Griot et al. 2025).

Finally, we address the topic deliberately skipped in Section 2.1.3: an early observation has been that LLMs performance appears to improve when prompted to "think step by step" (Kojima et al. 2022; Wei et al. 2022), a technique known as *Chain-of-Thought* (CoT) prompting, with one commenter characterizing this as working memory being an external organ for these systems (Newman 2023). Recent approaches automate CoT prompting (Zhang et al. 2023) or suggest generating CoT during training by manipulating training data (Gandhi et al. 2024; Lehnert et al. 2024) so users need not explicitly include special instructions in prompts. Industry has quickly adopted such ideas: as the time of writing, various state-of-the-art LLMs-based chatbots such as ChatGPT o3-pro, DeepSeek DeepThink (R1), and Gemini 2.0 Flash Thinking automatically generate an additional output akin to think-aloud protocols. One might interpret these as a foundational model thinking about and steering their own thinking process, even as traces of their metacognitive experiences in the sense investigated in this paper. However, recent studies raise doubts regarding the CoT's effectiveness in achieving reliable reasoning in foundational models (Stechly et al. 2024; Valmeekam et al. 2024). Whether such seemingly "think-aloud" CoTs should be interpreted as such has also been questioned, since they often are not valid traces of an algorithm (Kambhampati 2025). Kambhampati (2024) argued for interpreting chain-of-thought/reasoning traces less as (meta)cognitive functions and more as prompt augmentations for query answering systems. The questionable faithfulness of the generated CoT remains a key issue: has the LLM actually adhered to it, or simply hallucinated a CoT that sounds sensible? If performance were the only concern, it would not matter whether CoTs are actually metacognitive in nature or merely allow LLMs to better simulate the benefits of metacognition. However, when explainability is the focus, we must consider that LLM-presented train-of-thoughts may in fact be hallucinations that, instead of being truly transparent, may misleadingly feign human-like thinking. Unfortunately, evidence — beyond the Claude example (Lindsay et al. 2025) from our Introduction — is accumulating that the latter is indeed the case (Lanham et al. 2023; Turpin et al. 2023; Paul et al. 2024; Arcuschin et al. 2025). *Faithful* CoT generation now attracts attention by the research community but remains an unsolved problem.

# Declarations

**Data availability.** A companion website[25] offers interactive versions of several data visualizations in this paper to, e.g., filter by CMA. The collected raw data is available in a GitHub repository[26].

**Competing interests.** Nine authors (R.N., M.P., A.J., D.B., S.J., R.P., J.B., M.B., and R.M.) are current or former principle investigators, researchers or student employees of the Collaborative Research Center 1320 “EASE” at the University of Bremen (cf. the ‘Acknowledgments’ on page 1), which develops and evaluates the CRAM cognitive architecture. This review includes one of our authored CRAM papers (Nolte et al. 2023), resulting in a professional and reputational interest. We have no other conflicts of interest to declare that are relevant to the content of this article.

**Author Contribution.** R.N. conceived and designed the study, developed the methodology, and coordinated all phases of the systematic review. R.N. performed the majority of the literature screening and data extraction (as Screener A), conducted the complete synthesis and analysis, designed and implemented the accompanying website, conducted the search for related work, and wrote the manuscript draft (approximately 95% of the text). R.N. also updated and finalized the interactive data visualizations originally implemented by A.J.

M.P. contributed to refining the study scope and eligibility criteria, served as Screener B in the screening and data extraction process, contributed to the sections describing the purpose and algorithmic aspects of the reviewed systems, and critically reviewed and proofread the manuscript.

A.J. implemented the initial version of the interactive data visualizations used in the online companion website.

K.J. participated as Screener C during screening and data extraction and contributed to early discussions on the study’s scope. K.J. also provided critical proofreading of the final manuscript.

D.B. acted as Screener D in the screening and data extraction phase and was involved in shaping the early study scope.

S.J. contributed as Screener E and to the synthesis of the terminology section, compiling input data and drafting parts of the initial related text, while R.N. performed clustering and final integration.

R.P. provided advisory input throughout the project.

J.B., M.B., and R.M. are the principal investigators (PIs) of the overarching project that funded this research. They contributed to high-level conceptual oversight and provided management and institutional support. J.B. also proofread the manuscript.

[25] https://dml.uni-bremen.de/ease/review

[26] https://github.com/mrnolte/cma_review_data

# Index of Acronyms and Terms of Interest

*GPME* General Purpose Metacognition Engine (metacognitive architecture from the MCL Family) 25, 29–33, 36, 39

*H-CogAff* Human-CognitionAffect (metacognitive architecture) 18, 28, 32, 38 f., 42 f.

*H13* The metacognitive architecture introduced by Helgason (2013) 18, 31, 33 f., 38, 44

*HRI* Human-Robot-Interaction 50

*IDF* ArmarX Interpretable Data Format (see also ArmarX) 38

*IMXP* Introspective Meta-Explanation Pattern (see also Meta-AQUA, XP, and Meta XP) 41 f.

*ISAAC* Integrated Story Analysis And Comprehension (metacognitive architecture) 35, 37, 45, 48

*K10* The metacognitive architecture introduced by Kennedy (2010) 18, 26, 30, 38, 43

*K12* The metacognitive architecture introduced by Kennedy (2012) 18, 28, 32 ff., 38, 43

*K21* The metacognitive architecture introduced by Kennedy (2021) 18, 24, 32

*KnowRob* KnowRob (hybrid reasoning framework) 35

*KS02* The metacognitive architecture introduced by Kennedy and Sloman (2002) 18, 26, 32 ff., 36–40, 43

*LLM* Large Language Model 2 f., 9, 53

*MAMID* Methodology for Analysis and Modelling of Individual Differences (metacognitive architecture) 20, 28, 32, 38

*MCL* MetaCognitive Loop (metacognitive architecture from the MCL Family) 18, 28, 36 f., 48

*MCL Family* MetaCognitive Loop Family (Family of metacognitive architectures including MCL, MCL3, and GPME) 25 f., 28, 30, 38

*MCL3* MetaCognitive Loop Version 3 (metacognitive architecture from the MCL Family) 52

*Meta XP* Meta-Explanation Pattern (data structure; see also Meta-AQUA, XP, and IMXP) 25, 28

*Meta-AQUA* Meta-AQUA (metacognitive architecture; see also AQUA) 8 f., 16, 18, 20, 25 f., 28, 30–43, 45, 47 f.

*Metacat* Metacat (metacognitive architecture) 9, 16, 24 f., 28, 30 ff., 34, 36–40, 42, 44 f., 47 f., 53

*MIDCA* Metacognitive, Integrated Dual-Cycle Architecture (metacognitive architecture) 16, 18, 25–28, 31 f., 34–40, 45, 47 f.

*MISM* MISM (metamodel; metacognitive architecture) 18, 20, 25, 28, 30, 39 f., 48

*MT21* The metacognitive architecture introduced by van Ments and Treur (2021) 18, 29 f., 37 f., 44, 48

*NARS* Non-Axiomatic Reasoning System (metacognitive architecture) 32, 35, 37 f.

*NEEM* Narratively-Enabled Episodic Memory (data structure; see also CRAM) 35, 38

*OCC model of emotions* Ortony, Clore, and Collins (1988) model of emotions 24, 35 f.

*OMG* Object Management Group 4

*OWL* Web Ontology Language 35, 38

*PRISMA 2020* 2020 Preferred Reporting Items for Systematic reviews and Meta-Analyses 15

*PRODIGY* PRODIGY (metacognitive architecture) 3, 9, 16, 24, 30, 38, 44

*RCAA* Reflective-Cognitive Agent Architecture 28, 30 f., 33, 38, 42, 48

*REM* Reflective Evolutionary Mind (metacognitive architecture) 16, 28, 30, 38, 48

*SALS* Substrate for Accountable Layered Systems (metacognitive architecture) 26, 31 f., 34–39, 43, 48

*SCADS* Strategy Choice And Discovery Simulation (metacognitive architecture) 32, 36, 38 f., 43, 46 ff.

*Soar* Soar (metacognitive architecture) 7 f., 16, 18, 20, 24, 28, 32 ff., 36–40, 45, 49, 51

*SOMA* Socio-Physical Model of Activities (Ontology, see also CRAM) . . . . . . . . . 35, 48
*TalaMind* TalaMind (metacognitive architecture) . . . . . . . . . . . . . . . . . . . 27, 30, 38
*ToM* Theory of Mind . . . . . . . . . . . . . . . . . . . . . . . . . . . . . . . . . 16, 31, 39
*TOMASys* Teleological and Ontological Model of an Autonomous System (metacognitive architecture) . . . . . . . . . . . . . . . . . . . . . . . . . . . . . . . . . . . . . . . 7
*XAI* Explainable Artificial Intelligence . . . . . . . . . . . . . . . . . . . . . . . . . . 3, 53
*XP* Explanation Pattern (see also Meta-AQUA, IMXP, and Meta XP) . . . . . . . . . . 35, 41